\documentclass{aa}
\usepackage{graphicx,natbib,psfrag,amssymb,amsmath}
\usepackage{txfonts}
\usepackage{color}
\usepackage{textcomp}
\definecolor{noire}{rgb}{0,0,0} 
\definecolor{blue}{rgb}{0.1,0.5,0.7}
\bibpunct{(}{)}{;}{a}{}{,}

\usepackage{natbib}
\bibpunct{(}{)}{;}{a}{}{,}

\begin{document}

\title{Radio counterpart of the lensed submm emission in the cluster
MS0451.6$-$0305: new evidence for the merger scenario}

\subtitle{}

\author{A. Berciano Alba\inst{1,2}, L.V.E. Koopmans\inst{2}, M.A. Garrett \inst{1,3,4}, O. Wucknitz\inst{5} \&  M. Limousin\inst{6,7}}

\offprints{berciano@astro.rug.nl}

\institute{Netherlands Foundation for Research in Astronomy (ASTRON),
Postbus 2, 7990 AA Dwingeloo, The Netherlands \and Kapteyn
Astronomical Institute, University of Groningen, P.O. Box 800, 9700
AV, Groningen, The Netherlands \and Leiden Observatory, P.O. Box 9513,
2300 RA, Leiden, The Netherlands \and Centre for Astrophysics and
Supercomputing, Swinburne University of Technology, Mail number H39,
P.O. Box 218, Hawthorn, Victoria 3122, Australia \and
Argelander-Institut f\"ur Astronomie (AIfA), University of Bonn, Auf
dem H\"ugel 71, 53121 Bonn, Germany \and Laboratoire d'Astrophysique
de Marseille, UMR\,6110, CNRS-Universit\'e de Provence, 38 rue
Fr\'ed\'eric Joliot-Curie, 13\,388 Marseille Cedex 13, France \and
Dark Cosmology Centre, Niels Bohr Institute, University of Copenhagen,
Juliane Maries Vej 30, 2100 Copenhagen, Denmark}

\date{Received ...; accepted ...}

\abstract {SMM~J$04542-0301$ is an extended ($\sim 1 \arcmin$) submm
source located near the core of the cluster MS$0451.6-0305$. It has
been suggested that part of its emission arises from the interaction
between a LBG and two EROs at $\rm z\sim2.9$ that are multiply-imaged
in the optical/NIR observations. However, the dramatic resolution
difference between the sub-mm map and the optical/NIR images make it
difficult to confirm this hypothesis.}{In a previous paper, we
reported the detection of 1.4~GHz continuum radio emission coincident
with this sub-mm source using VLA archival data. To fully understand
the relation between this radio emission, the sub-mm emission, and the
optical/IR multiply-imaged sources, we have re-observed the cluster
with the VLA at higher resolution.}{The previous archival data has
been re-reduced and combined with the new observations to produced a
deep ($\sim$ 10~$\mu$Jy beam$^{-1}$), high resolution ($\sim
2\arcsec$) map centred on the cluster core. The strong lensing effect
in the radio data has been quantified by constructing a new lens model
of the cluster.}{From the high resolution map we have robustly
identified six radio sources located within SMM~J$04542-0301$. The
brightest and most extended of these sources (RJ) is located in the
middle of the sub-mm emission, and has no obvious counterpart in the
optical/NIR. Three other detections (E1, E2 and E3) seem to be
associated with the images of one of the EROs (B), although the NIR
and radio emission appear to originate at slightly different positions
in the source plane. The last two detections (CR1 and CR2), for which
no optical/NIR counterpart have been found, seem to constitute two
relatively compact emitting regions embedded in a $\sim 5\arcsec$
extended radio source located at the position of the sub-mm peak. The
presence of this extended component (which contributes 38\% of the
total radio flux in this region) can only be explained if it is being
produced by a lensed region of dust obscured star formation in the
center of the merger. A comparison between the radio and sub-mm data
at the same resolution suggests that E1, E2, E3, CR1 and CR2 are
associated with the sub-mm emission.}{The radio observations presented
in this paper provide strong observational evidence in favour of the
merger hypothesis. However, the question if RJ is also contributing to
the observed sub-mm emission remains open. These results illustrate
the promising prospects for radio interferometry and strong
gravitational lensing to study the internal structure of SMGs.}{}


\titlerunning{Internal radio structure of a faint sub-mm galaxy}
\authorrunning{Berciano Alba et al.}

\maketitle

\section{Introduction}

The detection of the cosmic infrared background (CIB) by the COBE
satellite \citep{PU96.1, HA98.1} established that about half of the
total radiation in the universe comes from dust-obscured galaxies that
are missing from optical surveys \citep[see][for a
review]{LA05.1}. This population of dusty objects was first
resolved by the IRAS and ISO satellites up to $z\sim1$, and turned out
to be dominated by luminous and ultra-luminous IR galaxies (LIRGs and
ULIRGs). The step into the high-z universe came with the advent of
sub-mm and mm surveys, since the far-infrared (FIR) luminosity peak of
high-z obscured galaxies is red-shifted into the sub-mm band
\citep{FR91.1,BL93.1}. As a result, sub-mm galaxies
(hereafter SMGs) turned out to be hundreds of times more numerous than
galaxies with similar luminosities in the local universe, suggesting
that they constitute the dominant contributor of the CIB and cosmic
star formation at $z\gtrsim1$. This illustrates the very important
role that SMGs play in the context of galaxy formation and evolution
\citep[see][for a review on (U)LIRGs and SMGs]{LO06.1}.

One decade after their discovery with SCUBA\footnote{Submillimeter
Common-User Bolometer Array, decommissioned in September 2005 from the
James Clerk Maxwell Telescope on Mauna Kea}
\citep{SM97.1,HU98.1,BA98.1,EA99.1}, it is generaly acepted
that SMGs are heavily dust-obscured galaxies at high redshift ($2<z
<3$) with ULIRG-like luminosities ($L_{\rm FIR} \sim 10^{12}~
L_\odot$) and star formation rates of the order of 1000
$M_\odot~yr^{-1}$. This enormous bolometric luminosity seems to be
dominated by star formation processes induced by galaxy
interactions/mergers, although a good fraction ($\sim 30-50 \%$) of
SMGs also host AGN activity \citep{AL05.1}. First estimates of their
physical properties indicate that SMGs are massive, gas rich systems
($M_{\rm gas}\sim10^{10}-10^{11}~M_\odot$) in which the starburst
region has a typical scale in the range $1-8$~kpc \citep{NE03.1,
CH04.1, GR05.1, TA06.1, WA07.1, BI08.1}. The available evidence also
suggests that SMGs might be the progenitors of massive local
ellipticals \citep{LI99.1, SM02.1, SM04.1, WE03.1, GE03.1,
AL03.1,AL05.1, SW06.1}.

However, these general properties of SMGs are based on the study of
the very brightest examples of this class of object ($\rm S_{850\mu
m}\gtrsim2$~mJy), and may not be representative of the entire SMG
population. In fact, according to \cite{KN08.1}, the dominant
contribution to the sub-mm extragalactic background comes from the
fainter (sub-)mJy sources that cannot usually be detected due to the
confusion noise of current instruments ($\rm S_{850\mu m}\sim
2$~mJy). The only way in which it has been possible to push bellow
this sensitivity limit, is by using the lensing magnification provided
by massive clusters of galaxies to increase the effective resolution
of SCUBA. This approach has improved the sensitivity of sub-mm maps by
factors of a few with respect to blank field surveys, although just a
handfull of faint SMGs have being identified so far \citep{SM02.1,
CO02.1, KN08.1}. Since so little is known about these intrinsically
faint sources, it is crucial to further investigate their properties
(e.g. spectral energy distributions, morphologies, redshifts, etc.)
and assess whether they are different from the observed properties of
brighter SMGs.

A promissing strategy to gather information about faint SMGs is to
study members that are multiply imaged by clusters of galaxies. In
this cases, the magnification factor can go up to 30 (or more),
providing not only the opportunity to detect but also to spatially
resolve the morphologies and internal dynamics of faint SMGs at a
level of detail far greater than would otherwise be possible
\citep[see][for an example of this technique in the
optical]{SW07.1}. To date, only one multiply-imaged faint SMGs has
been confirmed: SMM~J$16359+6612$, located near the core of the
cluster A2218 \citep{KN04.1,KN05.1,GA05.1,KN08.1,KN09.1}. There are,
however, two other clusters which seem to host multiply-imaged SMGs:
A1689 \citep{KN08.1} and MS$0451.6-0305$ \citep{CH02.1,
BO04.1,BE07.1}.

\begin{figure}
  \centering \includegraphics[width=9cm,angle=0]{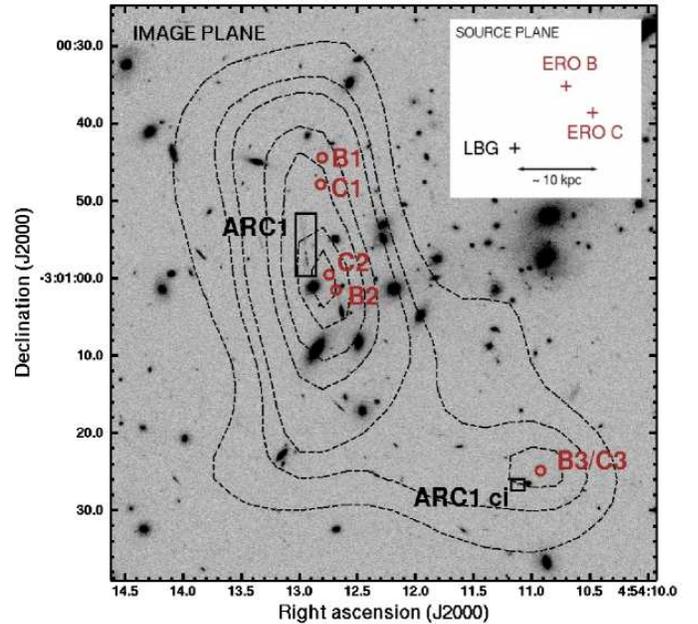}
  \caption{\textbf{Summary of the results presented in
\cite{BO04.1}.} SCUBA $850~\mu$m contour map of SMM~J$04542-0301$
superimposed upon a HST image of the center of the cluster
MS$0451.6-0305$. Highlighted with rectangles are an optical arc (ARC1)
and its counter image (ARC1 ci), produced by a LBG at $z_{\rm spect}=
2.911$. The red circles indicate the positions of five NIR sources
that have been interpreted as multiple images produced by two EROs
(ERO B, lensed as B1/B2/B3, and ERO C, lensed as C1/C2/C3) at $z_{\rm
model}=2.85 \pm 0.1$. Assuming $z=2.9$ for both the LBG and the EROs,
their predicted positions in the source plane would be located withing
a region of $\rm ~\sim 10~kpc$, suggesting that they constitute a
merger.}
  \label{intro}
\end{figure}

The case of MS$0451.6-0305$ is particularly interesting, since the
``sub-mm source'' (SMM~J$04542-0301$) is an elongated ($\sim 1
\arcmin$) region of $850~\mu$m emission which is coincident with an
optical arc and five NIR sources (see Fig. \ref{intro}).  While the
optical arc is the result of a strongly-lensed Lyman Break Galaxy
(LBG) at $z_{\rm spect}=2.911$, a lens model of the cluster predicts
that the set of NIR sources could be produced by two triply-imaged
EROs\footnote{Extremely Red Objects, photometrically defined as $R-K >
5.3$} located at almost the same redshift ($z_{\rm model}=2.85 \pm
0.1$). The model also indicates that the ERO pair and the LBG may
constitute a merger at $z=2.9$ in the source plane, with their
interaction likely being at the origin of the observed sub-mm emission
\citep[][B04 hereafter]{BO04.1}. Unfortunately, the low resolution
($\sim 15\arcsec$ at $850~\mu$m) and poor positional accuracy ($\sim
2\arcsec-3\arcsec$ rms) of SCUBA, makes it very difficult to confirm
the link between SMM~J$04542-0301$ and the proposed optical/NIR
sources. Moreover, the emission coming from the north-eastern and
central regions of the sub-mm emission cannot be reproduced by the arc
and the NIR sources (see Fig. 7 in B04), suggesting that it might
arise via other sources.

A possible way to overcome this resolution problem is by taking
advantage of the observed correlation between the radio synchrotron
and FIR emission in star-forming galaxies
\citep{VK73.1,CO82.1,HE85.1,GA02.1,AP04.1,BE06.1}, which also seems to
hold for SMGs out to $z\sim3$
\citep{KO06.1,VL07.1,IB08.1,MI09.1}. Thanks to this FIR-radio
correlation, radio interferometric observations can be used as a
high-resolution proxy for the rest-frame FIR emission observed in the
sub-mm. This approach has been extensively used to pinpoint the
position of SMGs in order to identify their faint optical counterparts
\citep[e.g.][]{IV00.1,BA00.1,CH05.1,IV07.1}.

In \cite{BE07.1}, we reported the detection of 1.4~GHz radio emission
coincident with SMM~J$04542-0301$ using VLA\footnote{Very Large Array
interferometer of the National Radio Astronomy Observatory (NRAO)}
archival data. Part of this radio emission is located in the region
between the optical arc and the ERO images, which is consistent with
the interacting region of the hypothetical merger being the source of
the observed radio and sub-mm emission. We also detected bright radio
emission in the central region of SMM~J$04542-0301$, although it is
not clear if this emission is produced by a high-z lensed object or
AGN activity associated with a cluster member.

In this paper, we present a higher resolution 1.4~GHz radio map of the
center of the cluster MS$0451.6-0305$ (MS0451 hereafter), obtained
after combining the previous VLA archival data (BnA configuration)
with new VLA high resolution observations (A configuration).  The
details of both sets of observations are presented in Sect. 2,
together with a description of the data analysis procedure, which has
been considerably improved with respect to \cite{BE07.1}. Section 3 is
dedicated to the analysis of the final combined data set, from which
we characterise the compact and extended radio emission detected
within SMM~J$04542-0301$. Identification of optical (HST/ACS F814W)
and NIR (SUBARU/CISCO K' band) counterparts for the radio detections,
as well as their connection with the sub-mm emission, is discussed in
Sect. 4. In Sect. 5 we describe a new lens model of the cluster
MS$0451.6-0305$, built to investigate the lensed nature of the compact
radio detections. A discussion about the merger scenario proposed by
\cite{BO04.1} including the results from the radio observations are
presented in Sect. 6. Summary and conclusions are presented in Sect. 7.

The adopted cosmology corresponds to a $\Lambda$CDM model with
$\Omega_{\rm m}=0.28$, $\Omega_{\lambda}=0.72$ and $h_{0}=73$
\citep{SP07.1}.

\section{Radio Observations and Data Reduction}  
\label{sec:reduction}

The 1.4~GHz observations of the cluster MS0451 presented in this paper
were obtained with the VLA in A and B configurations (A-array and
B-array hereafter) using the wide-field ``pseudo-continuum
mode''. This mode consists on two IFs of 25~MHz, centered at
1.3649~GHz and 1.4351~GHz. Each IF contains right and left circular
polarizations and it is split into seven 3.125~MHz channels. The
absolute flux density scale was set by the flux calibrator $0137+331$
(3C48), while $0503+020$ was used as the phase calibrator (PHCS
hereafter). The data reduction was conducted in classic \textsc{AIPS}
using our own semi-automatic pipeline written in Parseltongue
\citep[Python interface to classic \textsc{AIPS}, see][]{KE06.1}.

\subsection{A-array Observations}

The A-array observations consist on $2 \times 6$ hours of data
acquired the 5th and 10th of February 2006 (correlator integration
time of 3.3 seconds). Each IF of each observing day was independently
self-calibrated, resulting in four data sets that were in the end
combined to obtain the final A-array map of the target. In order to
avoid source smearing, no averaging in frequency nor time was applied
to the data. The different steps followed during the data reduction of
each epoch are described next.

First, we improved the positional accuracy of the antennas in our data
by applying the most recent baseline corrections determined by
NRAO. Corrupted data were visually inspected and flagged independently
for each channel, IF and polarization at different stages of the data
reduction process.

To set the flux scale, amplitude and phase corrections were calculated
for 3C48 using a model of this source provided by \textsc{AIPS}. Then,
3C48 was used to calibrate the bandpass shape. In the case of the
PHCS, we first derived a preliminary calibration in amplitude and
phase using a point source model. Subsequent self-calibration produced
a more accurate clean component model (CC model hereafter), which was
ultimately used to derive the final phase calibration.  After proper
interpolation, the resultant amplitude and phase solutions were
applied to the cluster data, and further refined through
self-calibration in order to produce the optimal map for our target.

\begin{figure*}[ht!]
  \centering
  \includegraphics[width=14cm,angle=0]{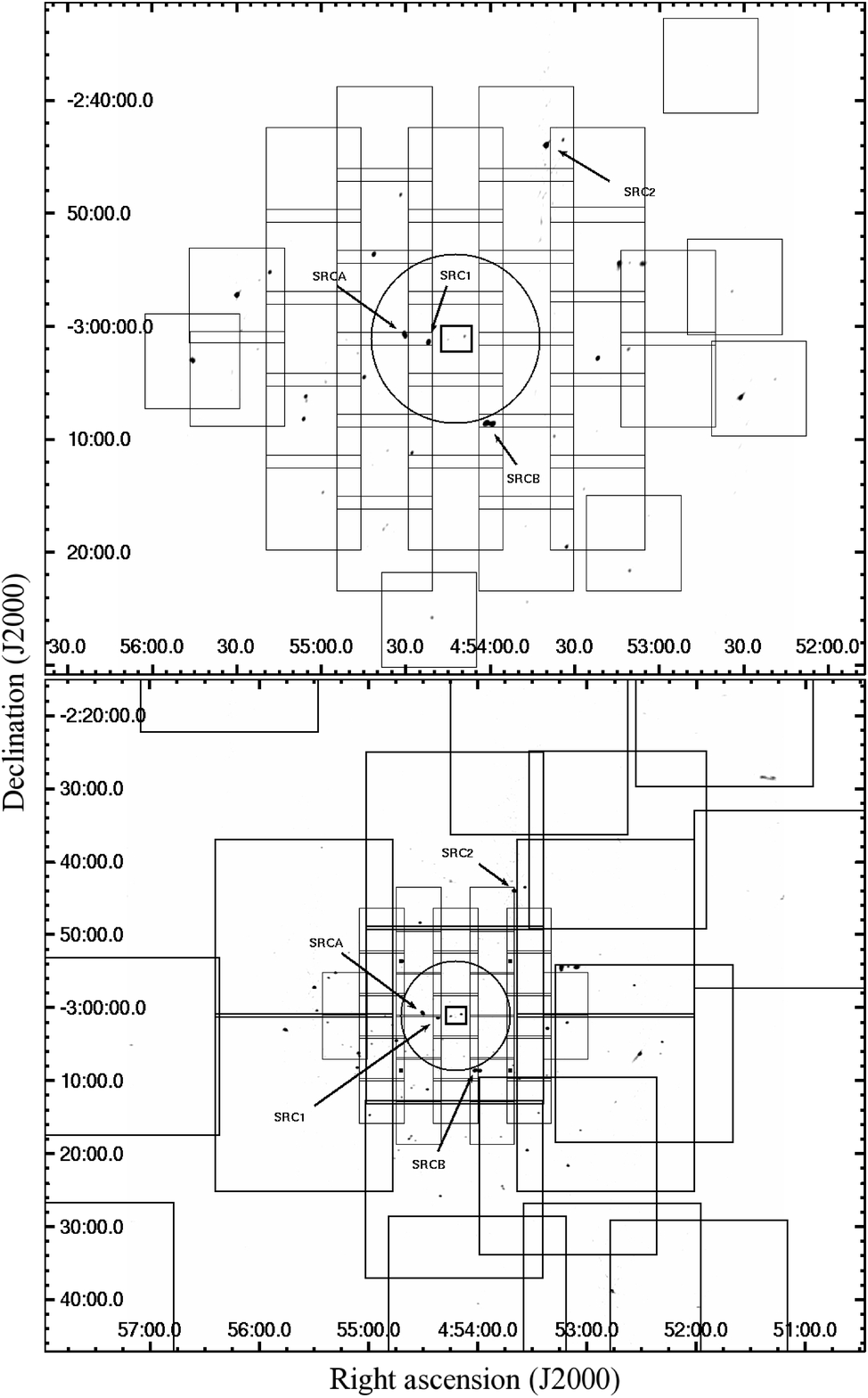} 
  \caption{VLA 1.4~GHz maps centered in the core of the cluster
  MS$0451.6-0305$. The maps were produced using the visibilities of
  the central 10~$\rm k \lambda$ region of the UV-plane only, to
  enhance the source visibility. The facets used during the
  selfcalibration of the cluster data are indicted with boxes, while
  the circle represents the VLA FWHM primary beam ($0.25^\circ$. at
  1.4~GHz). The central bold box corresponds to the target region
  mapped in Fig.\ref{A+Barray_R0.6}. The brightest sources in the
  field are indicated with labels. \textbf{Top Panel:} $1^\circ \times
  1^\circ$ map produced with the A-array observations. Each box is
  $\rm 8.4\arcmin \times 8.4\arcmin$. \textbf{Bottom Panel:}
  $1.5^\circ \times 1.5^\circ$ map produced with the B-array
  observations. The big boxes (used in the first step of the
  selfcalibration procedure) are $\rm 27.8\arcmin \times 27.8\arcmin$
  in size, whereas the small boxes (used in the second step) are
  $6.8\arcmin \times 6.8\arcmin$.}

  \label{fields}
\end{figure*}

Figure \ref{fields} (top) shows a low resolution $1 \times 1$~deg map
of the cluster produced with the inner 10~k$\lambda$ baselines of the
A-array data. Due to the significant number of bright sources located
outside the main lobe of the primary beam (black circle), 37 facets
were used for the imaging part of the self-calibration process: 31
overlapping facets that cover the central 0.32~deg radius region, and
6 individual facets centered at the positions of bright NVSS
sources\footnote{ The NRAO VLA Sky Survey (NVSS) is a 1.4~GHz
continuum survey covering the entire sky north of
$-40$~deg. declination, with a completeness limit about 2.5~mJy}
located in an annulus between 0.32 and 0.5~deg. radius. The size of
each facet is $2048 \times 2048$~pixels, with a pixel scale of
0.247\arcsec. Areas for {\scriptsize CLEANING} were restricted by
placing boxes around the brightest sources in each facet. The
weighting scheme used for imaging was selected by the {\scriptsize
IMAGR} parameter {\scriptsize ROBUST} (R hereafter), with value
ranging from $+5$ (pure natural weighting) to $-5$ (pure uniform
weighting).  To produce a good model of the compact emission, $\rm
R=0$ was used until the last iteration of the self-calibration
procedure, where we switched to $\rm R=5$ to include the extended
emission.

From the sources indicated in the top panel of Fig. \ref{fields}, SRC1
is compact, and it is the brightest and closest to the target, so it
should provide the best self-calibration solutions in the target
region. However, the extended sources SRCA and SRCB dominate the
solutions in the short baselines. In order to get the final
calibration only from SRC1 for all baselines, all the sources located
outside the central facet were subtracted from the data using the
final CC model of the outer 37 facets. Finally, the resultant data set
was self-calibrated again using $\rm R=5$ during the imaging process.

This procedure was followed for each IF of each epoch, resulting in
four independently calibrated data sets (in which only the sources
from the central facet remain) that were concatenated with
{\scriptsize DBCON} to produce the final A-array map of the target.

\subsection{B-array Observations}

The B-array observations, made the 9th and 10th of June 2002, were
retrieved from the NRAO data archive system
\footnote{Project ID AN0109, PI: Nakanishi} (7.8 hours in total,
correlator integration time of 10 seconds). The first time that these
observations were analyzed \citep{BE07.1}, they were treated as a
single epoch and averaged on time, with no independent calibration for
each IF. However, since our goal is to combine the data from both
array configurations to produce a more sensitive map of the target,
the B-array observations were re-reduced following essentially the
same procedure described in the previous section. The only difference
corresponds to the self-calibration of the cluster data, which we now
describe in detail.

\begin{table*}
\centering
\caption{ \textbf{Information about the identifyed radio sources in
 the core of MS0451}. The columns show: position (RA, DEC), peak flux
 density ($S_{\rm Pk}$), total flux density inside the 2~$\sigma$
 contour ($S_{\rm 2\sigma}$), total flux density obtained by adding
 the flux from all the clean components used to model the source (
 $S_{\rm CC}$), total flux density inside the 2~$\sigma$ contour in
 the residual map ($S_{\rm resid}$), comparison between the total flux
 derived from the CC model and the $2~\sigma$ contours $\left(
 \frac{S_{\rm CC} - S_{2\sigma}}{S_{2\sigma}} \right)$, comparison
 between the total flux derived from the CC model and the gaussian fit
 $\left( \frac{S_{\rm CC} - S_{\rm Gauss}}{S_{\rm Gauss}} \right)$ and
 signal-to-noise ratio (SNR). The quoted errors in RA and DEC
 correspond to the expected positional uncertainty of a point-like
 source in the map ($\rm \emph{FWHM} / 2 \times \emph{SNR}$, were
 \emph{FWHM} is the full with half maximum of the clean beam), while
 the error in the flux peak is estimated to be the same as the rms
 noise of the map. The error in the $2\sigma$ flux was calculated as
 $rms \times \sqrt{N}$, where \emph{rms} is the rms noise of the map,
 and \emph{N} is the number of beams within the area delimited by the
 $2\sigma$ contour of the source. The raw ``RJ bright'' lists the
 fluxes for RJ estimated withouth taking into account the $4\sigma$
 west extension. Note that the positions listed here were derived
 before aligning the radio map respect to the HST image.}
\begin{tabular}{lcccccccccc}
\hline\hline 
\noalign{\smallskip}
Name & RA ($+4^{h}$ $54^{m}$) & DEC ($- 3^{\circ}$)         & $S_{\rm Pk}$      & $S_{2\sigma}$ & $S_{\rm CC}$ & $S_{\rm resid}$ & $\frac{S_{\rm CC} - S_{2\sigma}}{S_{2\sigma}}$   & $\frac{S_{\rm CC} - S_{\rm Gauss}}{S_{\rm Gauss}}$  & SNR \\ 
                    & J2000 (sec)            & J2000 ($\arcmin$, $\arcsec$)& ($\rm \mu Jy~beam^{-1}$) & ($\mu$Jy)  & ($\mu$Jy)   & ($\mu$Jy)  &  ($\%$)                              &   ($\%$)                        &     &      \\
\noalign{\smallskip}
\hline 
\noalign{\smallskip}
E1        & $12.78 \pm 0.01$ & $00:43.8 \pm 0.2$ & $66\pm10$ & $42.5 \pm12$ & 61.6  & -1.4  & 45  &  17  & 6.4    \\
CR1       & $12.88 \pm 0.01$ & $00:53.5 \pm 0.2$ & $47\pm10$ & $36.6 \pm12$ & 47.2  & 2.9   & 29  & -16  & 4.6   \\
CR2       & $12.86 \pm 0.01$ & $00:59.3 \pm 0.2$ & $47\pm10$ & $43.8 \pm12$ & 60.1  & -0.1 & 37  & -21  & 4.6   \\
E2        & $12.66 \pm 0.01$ & $01:01.2 \pm 0.2$ & $51\pm10$ & $43.2 \pm12$ & 52.0  & 5.1   & 20  & -15  & 5.0   \\
RJ bright & $12.54 \pm 0.01$ & $01:17.2 \pm 0.1$ & $92\pm10$ & $121.2\pm17$ & 135.9 & --    & 12  & -17  & 8.9   \\
RJ        & $12.54 \pm 0.01$ & $01:17.2 \pm 0.1$ & $92\pm10$ & $168.0\pm22$ & 195.3 & 10.5  & 16  & --   & 8.9   \\ 
E3        & $10.97 \pm 0.01$ & $01:24.8 \pm 0.2$ & $54\pm10$ & $63.3 \pm15$ & 66.8  & 6.5  & 5   & -28  & 5.2    \\		  
\noalign{\smallskip}                                                                         
\hline		  
\noalign{\smallskip}		  
Fd    & $15.162  \pm 0.003  $ & $01:10.83 \pm 0.05$ & $199 \pm10$  & $1180.9\pm42 $ & 1260  & 29.5   & 7  & --  & 19.2      \\
Fe    & $09.2733 \pm 0.0007 $ & $00:50.90 \pm 0.01$ & $1068\pm10$  & $1769.2\pm34 $ & 1870  & 16.3   & 6  & --  & 103.3     \\
\hline

\end{tabular}
\label{tab_maxfit_R0.6}
\end{table*}

Figure \ref{fields} (bottom) shows a 1.5 $\times$ 1.5 deg low
resolution map of the cluster produced with the inner 10~K$\lambda$
baselines of the B-array data. Note that SRC2 is brighter than SRC1,
contrary to the situation in the A-array data (top) where SRC2 is
resolved. However, since SRC2 is considerably further away from the
phase center, the best self-calibration solutions on the target region
will still be provided by SRC1. Therefore, the data was
self-calibrated in three steps. First, a set of 7 overlapping facets
was used to image the central 0.4 deg radius region of the data, while
another 36 facets were centered at the positions of NVSS sources
located in an annulus between 0.4 and 1.6 deg radius (big boxes in the
bottom panel of Fig. \ref{fields}). Each facet is $2048 \times
2048$~pixels, with a pixels size of $0.814\arcsec$. Once the
self-calibration procedure was completed, the CC model from the 42
outer facets was subtracted from the data, leaving only the sources
located in the central facet. In the second step, the central region
of the resultant data set is imaged again using 31 overlapping facets
of $500 \times 500$~pixels. After following the same self-calibration
procedure used in the first step, the best CC model of the outer 30
facets was subtracted from the data. In the third step, this final
data set is calibrated (phase only) using a solution interval of 5~min
and $\rm R=5$.

As in the case of the A-array observations, the four independently
calibrated data sets obtained in this way (one per epoch and IF) were
then combined and imaged to produce the final B-array map of the
target.

\section{Analysis of the A+B-array map}  
\label{sec:analisis}

Before the A-array and B-array observations could be combined to
produce the final radio map of the cluster center, it was necessary to
correct for a single position shift between the two calibrated
data-sets. This correction was achieved by performing a global
phase-selfcalibration of the B-array data set using the best A-array
CC model.

\begin{figure}
\begin{tabular}{c}
\includegraphics[width=8cm,angle=0]{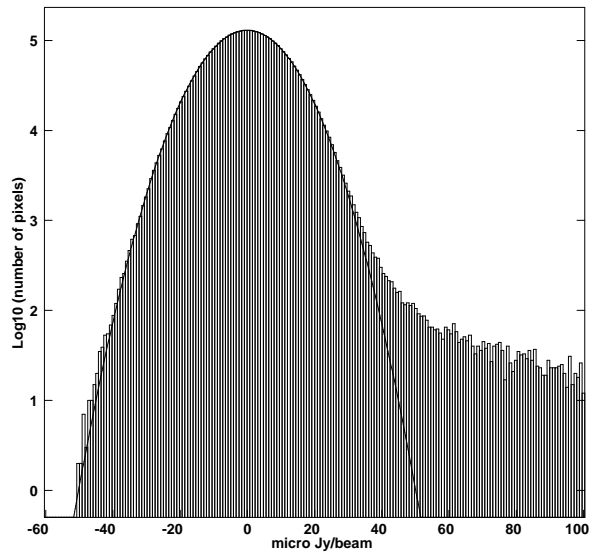} 
\end{tabular}
\caption{Histogram of the A+B array clean map produced using
$\rm R=0.6$. The black curve is the best Gaussian fit to the noise peak,
while the positive excess on the right side is due to source
emission. The good agreement between the fit and the data in the
negative side of the gaussian indicates that the noise of the map is
well behaved, making faint detections more reliable.}

\label{hist}
\end{figure}

To produce the map with the best compromise between sensitivity and
noise properties, we explored the full range of values available for
the {\scriptsize IMAGR} parameter {\scriptsize ROBUST}. The histograms
of the resultant clean maps show that the noise distribution is
closest to Gaussian when $\rm R=0.6$ (see Fig. \ref{hist}). Note also
that the rms noise of the $\rm R=0.6$ map ($\sim 10~\mu$Jy) is only
$2\%$ higher than the one obtained for $\rm R=5$, which is the most
sensitive map that can be produced with this data.

\subsection{Catalog of radio sources}
\label{sec:radio_catalog}

\begin{figure*}[t!]
 \centering
 \includegraphics[width=17cm,angle=0]{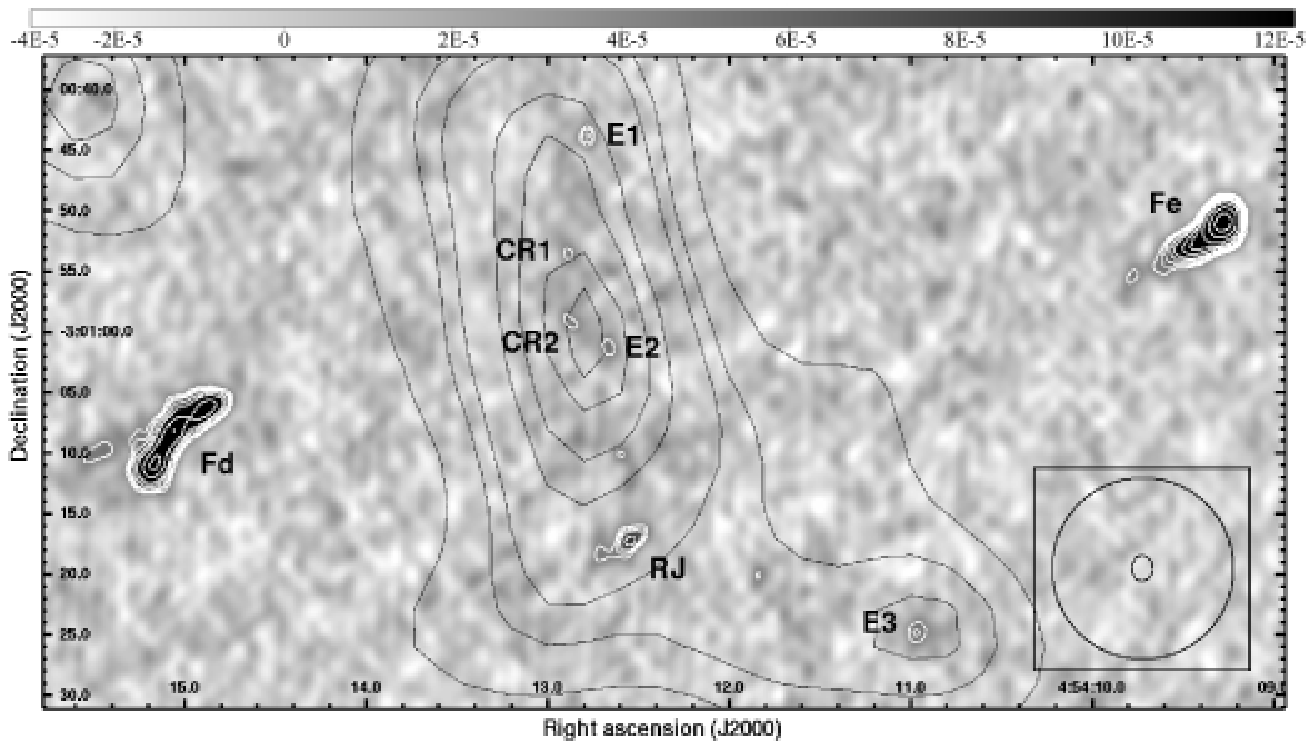}

 \caption{VLA 1.4~GHz map derived from the A+B-array data set using
$\rm R=0.6$. The greyscale has units of Jy beam$^{-1}$, and the
corresponding contours (white) are drawn at -4, 4, 5, 6, 8, 10, 14,
18, 28, 48 $\&$ 78 times the 1$\sigma$ noise level of 10.34~$\mu$Jy
beam$^{-1}$. The black lines (4, 6, 7, 9, 10, 11 $\&$ 11.5 mJy
beam$^-1$) correspond to the SCUBA 850~$\mu$m contour map published in
\cite{BO04.1}. A comparison of the beam size of both contour maps is
shown in the bottom right corner: $2.14\arcsec \times 1.71\arcsec$ at
a position angle of $-1^{\circ}.43$ for the radio map, and $15\arcsec
\times 15\arcsec$ for the submm map.}
\label{A+Barray_R0.6}

\centering
\includegraphics[width=17cm,angle=0]{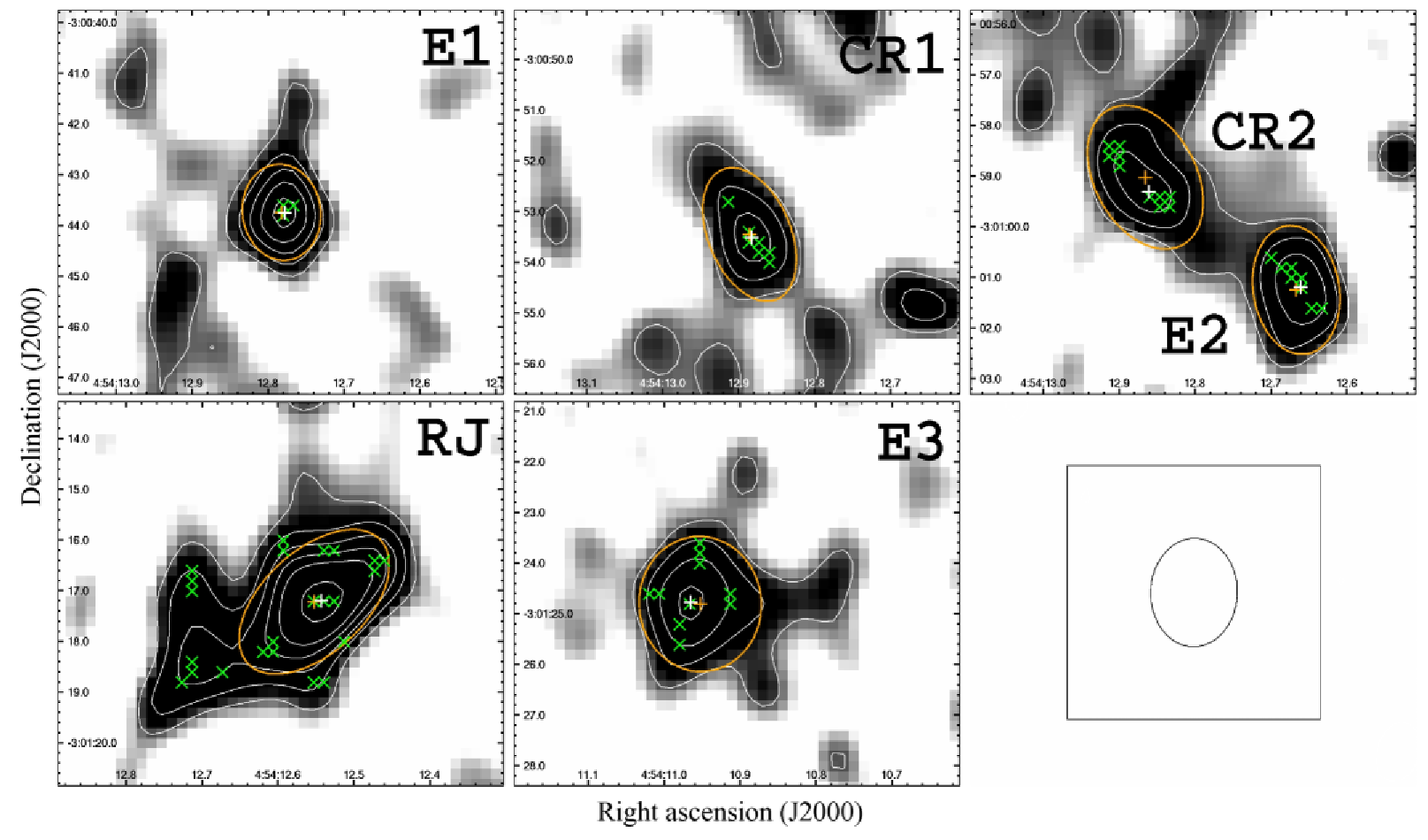} 

\caption{Detail of the six radio sources located within the sub-mm
emission region in Fig. \ref{A+Barray_R0.6}. The white radio contours
are drawn at 2, 3, 4, 5, 6 $\&$ 8 times the 1$\sigma$ noise level of
10.34~$\mu$Jy beam$^{-1}$. The radio peak positions ($\rm S_{Pk}$)
listed in Table \ref{tab_maxfit_R0.6} are indicated with white
plus-symbols. The green crosses correspond to the positions of the
clean components in each source model. Orange plus-symbols and
ellipses represent the center and FWHM of the best fitting Gaussian
for each source. A scaled version of the radio beam ($2.14\arcsec
\times 1.71\arcsec$ at a position angle of $-1^{\circ}.43$) is
included in the bottom right panel.}

\label{gaussfit_R0.6}
\end{figure*}

Figure \ref{A+Barray_R0.6} shows a detail on the $\rm R=0.6$ map of the
cluster center produced by cleaning the data down to a depth of $5.5~
\mu$Jy. The contours indicate the region of extended sub-mm emission
reported by B04, in which six of the eight identified radio sources
are located. Postage stamps of these six sources and their CC models
are shown in Fig. \ref{gaussfit_R0.6}.

\begin{table*}
\centering
\caption{ \textbf{Gaussian fitting results for the identified compact
 radio sources in the core of MS0451}. The columns show:
 position (RA, DEC), peak flux density ($S_{\rm Pk}$), total flux
 density ($S_{\rm Gauss}$), deconvolved Gaussian sizes (major axis,
 minor axis and position angle) with their corresponding formal
 errors, and the sum of the square of the residuals reported from the
 fit (SSR). Cases where the parameters of the Gaussian fits are not
 well constrained are indicated by a dash. Note that the positions
 listed here were derived before aligning the radio map respect to the
 HST image.}
\begin{tabular}{lcccccccc}
\hline\hline
\noalign{\smallskip}
 Name & RA ($+4^{h}$ $54^{m}$) & DEC ($- 3^{\circ}$)  & $S_{\rm Pk}$ & $S_{\rm Gauss}$ & Maj Axis & Min Axis & PA & SSR                \\ 
      & J2000 (sec) & J2000 ($\arcmin$, $\arcsec$)& ($\rm \mu Jy~beam^{-1}$) & ($\mu$Jy) & (arcsec) & (arcsec) & (deg) & ($\rm \times 10^{-9}~Jy^{2}~beam^{-2}$)\\
\noalign{\smallskip}
\hline 
\noalign{\smallskip}
E1         & $12.782\pm 0.007$  & $00:43.7\pm0.1$ & $66.1\pm10.3$ & $53.0  \pm15.5$ & $0.00  \pm1.14$   & --               & --              & 2.2 \\
CR1        & $12.89\pm  0.01$   & $00:53.5\pm0.3$ & $45.6\pm10.2$ & $56.2  \pm20.4$ & $1.87  \pm1.87$   & $0.00  \pm1.17$    & $31.4  \pm44.7$   & 0.6 \\
CR2        & $12.87\pm  0.01$   & $00:59.0\pm0.3$ & $45.9\pm10.0$ & $76.3  \pm24.5$ & $2.28  \pm1.22$   & $0.73  \pm0.99$    & $37.6  \pm3.8$    & 1.8 \\
E2         & $12.67\pm  0.01$   & $01:01.2\pm0.2$ & $51.6\pm10.3$ & $61.2  \pm20.0$ & $1.44  \pm1.44$   & $0.00  \pm1.07$    & --              & 0.9 \\
RJ bright  & $12.55\pm  0.01$   & $01:17.2\pm0.1$ & $82.9\pm9.8$  & $164   \pm27.6$ & $3.02  \pm0.56$   & $0.51  \pm0.73$    & $128   \pm10.9$   & 12.8\\
E3         & $10.95\pm  0.01$   & $01:24.8\pm0.2$ & $52.6\pm9.9$  & $92.5  \pm25.5$ & $1.71  \pm0.95$   & $1.58  \pm1.58$    & $70.5  \pm39.0$   & 3.4 \\            
\hline \\

\end{tabular}
\label{tab_gaussfit_R0.6}
\end{table*}

The radio detections located outside the sub-mm emission, labeled Fd
and Fe, are the same sources reported in \cite{BE07.1}. Their
morphology in the new high resolution radio map is very elongated,
typical of radio jets produced by AGNs. In the case of the other 6
sources, the morphology and distribution of their clean components
show that, with the exception of the (almost) point-like structure of
E1, the rest of the sources are extended. In particular, RJ shows a
bright elongated main body and an irregular $4\sigma$ extension
towards the west. Note however, that the amorphous extensions present
in the $2\sigma$ contours of E1, CR2 and E3 are not considered to be
part of their real structure. In the following, references to the
$2\sigma$ contours of these sources will implicitly exclude these
amorphous extensions.

The peak positions and peak fluxes of all the radio detections were
calculated with the task {\scriptsize MAXFIT}, which fits a quadratic
function to a rectangular area that encloses the source (in our case,
a rectangle delimited by the $2\sigma$ contours of the source). The
total flux, on the other hand, was determined in two different ways:
(i) by integrating the emission over the area defined by the $2\sigma$
contour of the source ($S_{2\sigma}$), and (ii) by adding the flux of
all the clean components of the source model ($\rm S_{\rm CC}$). In
this way, we can calculate the flux excess contained in the clean
component model respect to the $2\sigma$ flux, which is an estimate of
the amount of flux contained in the wings of the PSF-convolved sources
below the $2\sigma$ contour.

\begin{figure}
  \centering
  \includegraphics[width=6cm,angle=0]{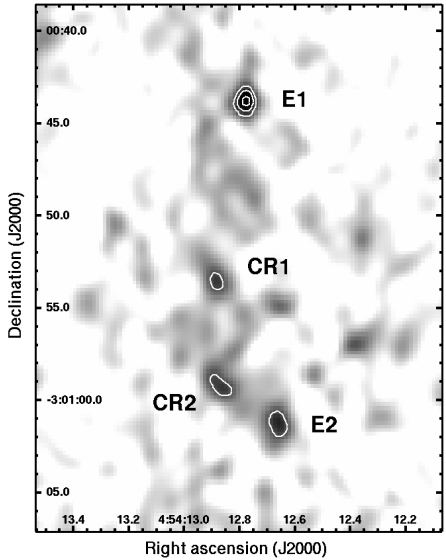}
  \caption{Detail of Fig. \ref{A+Barray_R0.6}. The grey scale has been
  modified to make more evident the low level background in this
  region of the radio map.}
  \label{merger_zoom}
\end{figure}

The results obtained for the different parameters are listed in Table
\ref{tab_maxfit_R0.6}. Note that, since the cleaning procedure is
stopped at certain flux level (half a sigma below the noise level), it
is expected to have some left-over emission from the sources in the
residual map, specially in the case of extended sources. To estimate
this left-over flux, we have integrated the emission inside the
$2\sigma$ contours of each source in the residual map (see $S_{\rm
resid}$ in Table \ref{tab_maxfit_R0.6}). The results show that the
$S_{\rm CC}$ for CR1 and RJ is underestimated by $\sim 6\%$, and by
$\sim 10\%$ in the case of E2 and E3. The negative values for $S_{\rm
resid}$ obtained for E1 and CR2 indicate that all the flux has been
included in the CC model.

We also tried to characterize the properties of the six sources
located within the sub-mm emission region by comparing them with
Gaussians. The task {\scriptsize JMFIT} was used to fit simultaneously
all the parameters of an elliptical Gaussian in the same rectangular
area used with {\scriptsize MAXFIT}. Since the shape of RJ is far from
being elliptical , we restricted the Gaussian fitting to the area that
only contains the $2\sigma$ contours of the bright main body. The
center and FWHM of the best fitting gaussians are shown in
Fig. \ref{gaussfit_R0.6}.

The parameters derived from the best fits and their formal errors are
listed in Table \ref{tab_gaussfit_R0.6}, where the sum of the square
of the residuals is a measure of the relative goodness of the fit. As
expected, the worst fits are obtained for RJ and E3, since they have
the most irregular shapes.  The best fits correspond to CR1 and E2,
but note that the FWHM of the Gaussian in CR1 does not include the
north-east $2\sigma$ extension.

When compared with Table \ref{tab_maxfit_R0.6}, we see that the
Gaussian fit provides basically the same estimates for the position
(RA, DEC) and flux peak ($S_{\rm Pk}$). The nominal integrated flux
($S_{\rm Gauss}$) is in general larger than $S_{\rm CC}$ (except for
E1), but consistent within the errors. The largest discrepancies are
found for E3 and CR2 (see $\frac{S_{\rm CC} - S_{\rm Gauss}}{S_{\rm
Gauss}}$ in Table \ref{tab_maxfit_R0.6}), probably because $S_{\rm
Gauss}$ is including some flux from the $2\sigma$ amorphous extensions
in these two sources that were not accounted for in the clean
component model. 

Finally, we would like to emphasize that, despite the faintness of CR1
and CR2 ($\rm SNR \sim 4.6$), the histogram presented in
Fig. \ref{hist} and the distribution of $-4\sigma$ peaks in the final
map (only a small number, located far away from the target region)
suggests that they are real detections.

\subsection{Disentangling the compact and extended radio emission}
\label{sec:compact_vs_extended}

\begin{figure*}
 \centering 
 \includegraphics[width=15cm,angle=0]{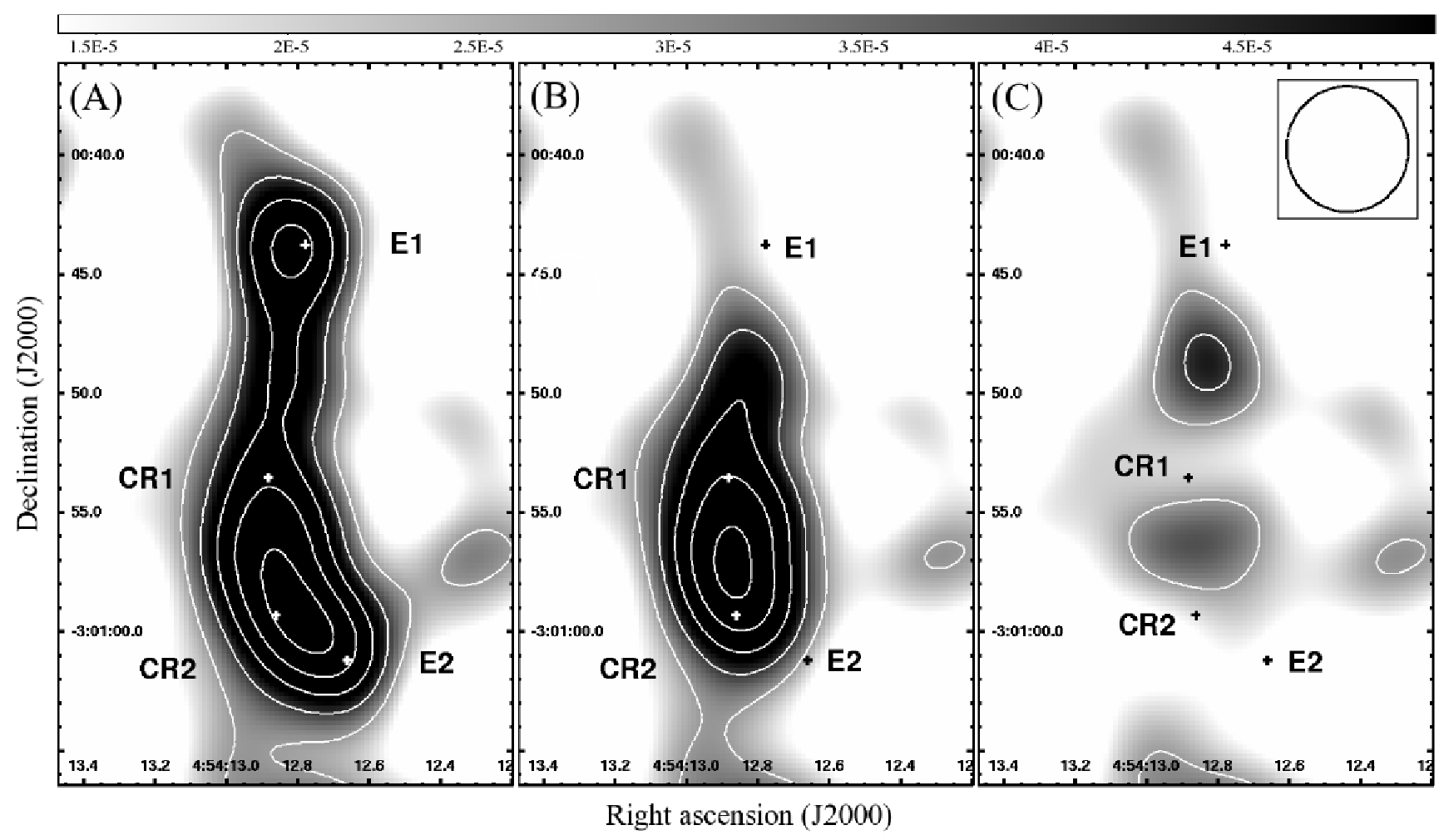}
 \caption{VLA 1.4~GHz maps obtained by tapering the A+B-array data set
 with a Gaussian of $\rm FWHM=30~k\lambda$. The crosses indicate the
 positions of the radio detections identified in
 Fig. \ref{A+Barray_R0.6}. Panel (A) includes all the emission between
 E1 and E2. In panel (B) the CC models of E1 and E2 were subtracted
 from the data before the tapering and cleaning procedure. Panel (C)
 shows what is left when CR1 and CR2 are also subtracted from the
 data. The gray-scale has units of Jy beam$^{-1}$, and the
 corresponding contours are drawn at 2,3,4,5 $\&$ 6 times the
 1$\sigma$ noise level of 14 $\mu$Jy beam$^{-1}$. The dimensions
 of the beam are $5.27\arcsec \times 5.12\arcsec$ at a position angle
 of $-88^{\circ}.25$ (panel (C), top right corner).}.
 \label{tap30_comparison}
\end{figure*}

As can be seen in Fig. \ref{merger_zoom}, the region between E1 and E2
shows an enhanced low level background when compared with the rest of
the map, suggesting the presence of diffuse radio emission that has
been partially resolved out. From now on, we will refer to this
diffuse emission as the \emph{extended component}, whereas E1, E2, E3,
CR1 and CR2 will be consider \emph{compact sources} by comparison.

If present, this extended component would only be detectable on short
baselines, while the emission produced by compact sources will be
picked up by all baselines at different resolution levels. Therefore,
a map produced using only the short baselines does not allow us to
distinguish between extended emission and a compact source observed at
low resolution. The only way to isolate the extended component is by
subtracting a model of the compact sources from the data.

Given the faintness of CR1 and CR2, and the complicated low level
background in this region, the choice of boxes during the cleaning
procedure becomes a highly subjective issue, and even the decision of
cleaning CR1 and CR2 (and hence their CC models) can be
questionable. For this reason, we produced a new version of the map
presented in Fig. \ref{A+Barray_R0.6} (following the same procedure)
in which CR1 and CR2 have not been cleaned. Then, we calculated
$S_{\rm Pk}$, $ S_{2\sigma}$ and $S_{\rm CC}$ for all the sources in
the new map, and compared them with the values presented in Table
\ref{tab_maxfit_R0.6}. In the case of E2 (which is the closest source
to CR1 and CR2) we find differences of $\sim 10\%$ for $S_{\rm CC}$
and $\sim 14\%$ for $\rm S_{2\sigma}$, whereas for the rest of the
sources the differences in all three parameters are less than
$7\%$. Since these variations are well withing the errors estimated
for $S_{\rm Pk}$ and $S_{2\sigma}$ in Table \ref{tab_maxfit_R0.6}, we
can use the CC models of CR1 and CR2 presented in
Fig.\ref{gaussfit_R0.6} without compromising the results.

Fig. \ref{tap30_comparison} shows three $\sim 5\arcsec$ resolution
maps of the region between E1 and E2, produced by tapering the data
with a Gaussian of $\rm FHWM=30~k\lambda$. In map (A) no compact
sources were subtracted before the tapering and imaging. The central
map (B) shows the result of subtracting the CC model of E1 and E2 from
the data, and map (C) shows what is left after subtracting all the
compact sources. Given the robust detection of a $6.6\sigma$ elongated
source in panel (B), the presence of resolved extended radio emission
in this region of the $\sim 2\arcsec$ map (Fig. \ref{merger_zoom}) is
confirmed. Using the $2\sigma$ contour of map (B) as template, the
integrated fluxes in panels (B) and (C) are $S_{\rm B}=173.52 \pm
\mu$Jy and $S_{\rm C}=84.14 \pm \mu$Jy. The flux error was calculated
as $rms \times \sqrt{N}$, where \emph{rms} is the rms noise of the
map, and \emph{N} is the number of beams within the area delimited by
the $2\sigma$ contour of map (B). These numbers indicate that $\sim
50\%$ of the flux in panel (B) has been included in the $\sim
2\arcsec$ CC model of CR1 and CR2. Therefore, it seems that CR1 and
CR2 constitute two relatively compact regions of an extended radio
source located between E1 and E2. However, given the faintness of CR1,
CR2 and the extended component in panel (C), deeper observations will
be required to confirm this result.

Finally, to estimate the relative contribution of all the compact
sources respect to the extended emission, we used the $2\sigma$
contour of map (A) as template to calculate the integrated flux of
each map in Fig. \ref{tap30_comparison}: $S_{\rm A}=285 \pm 33~\mu$Jy,
$S_{\rm B}=204 \pm 33~\mu$Jy and $S_{\rm C}=109 \pm 32~\mu$Jy. The
comparison of these fluxes indicates that $38\%$ of the emission comes
from an extended component, whereas $62\%$ is produced by the compact
sources ($28\%$ from E1 and E2, $34\%$ from CR1 and CR2). As it will
be discussed in detail in Sect. \ref{radio_vs_optical}, the presence
of this extended radio component constitutes a strong observational
evidence in favor of the merger scenario proposed in B04.

\section{Multiwavelength Counterparts} 
\label{subsec:look-optic-count}

\begin{table*}
\centering
\caption{\textbf{Observations details.} }
\begin{tabular}{lcccccc}
\hline\hline
\noalign{\smallskip}
               Instrument    & Frequency (filter) & Observing dates  & Total exposure time  & depth              & resolution  & references   \\                             
                             &       ($\mu$m)     &   (dd/mm/yy)     & (ks)                 & (magnitudes)       & (arcsecs)   &               \\
\noalign{\smallskip}
\hline 
\noalign{\smallskip}				 								
               HST/ACS       & 0.83 (F814W)       &     $-$          & 2.4-2.7 per pointing &      25            &   1.0       &    \cite{MO07.1}$^{1}$\\ 
                             &                    &                  &                      &                    &             &    Smith et al., in prep.        \\
	       SUBARU/CISCO  & 2.13   (K')        &  17-18/11/2000   &     9.22             &      22.1          &   0.6-0.8   &    \cite{TA03.1}        \\
               JCMT/SCUBA    & 850                &  03/09/1998      &     22               &      $-$           &   15.0      &    \cite{BO04.1}        \\ 
\hline
\noalign{\smallskip}
\multicolumn{7}{l}{$^1$ Data products from their wide field survey in MS0451 are available at http://www.astro.caltech.edu/~smm/clusters/}\\
\multicolumn{7}{l}{Further information on the data and survey can be found in Chapter 2 of the PhD thesis of Sean Moran.}\\

\end{tabular}
\label{multi_data}
\end{table*}

Table \ref{multi_data} summarizes the relevant information about all
the multiwavelength data of MS0451 (optical, NIR and sub-mm) that we
have collected from the literature to compare with our radio map. To
identify possible counterparts of the radio sources, the radio and K'
band observations were aligned with respect to the HST
image\footnote{astrometry matched to Subaru/Suprime-Cam observations
of MS0451 for the PISCES survey \citep{KO05.1}} using the
{\scriptsize IRAF} package {\scriptsize CCMAP}. For the alignment of
the radio map we used 13 compact radio sources with reliable bright
counterparts in the HST image, whereas 93 compact NIR sources were
used to align the K' band image. We find an rms scatter of
$0.29\arcsec$ between the radio map and the HST image, and
$0.06\arcsec$ between the K' band and the HST images. Therefore, the
inferred error in the alignment is $0.29/ \sqrt{13}=0.08\arcsec$ for
the radio map and $0.06/ \sqrt{93} =0.006\arcsec$ for the K' band
image. These errors are negligible compared with the (relative)
astrometric accuracy of the sources in each map, so they will be
ignored in further analysis.

\subsection{Comparison between radio and optical/NIR}
\label{radio_vs_optical}

\begin{figure*}
  \centering
  \includegraphics[width=18cm,angle=0]{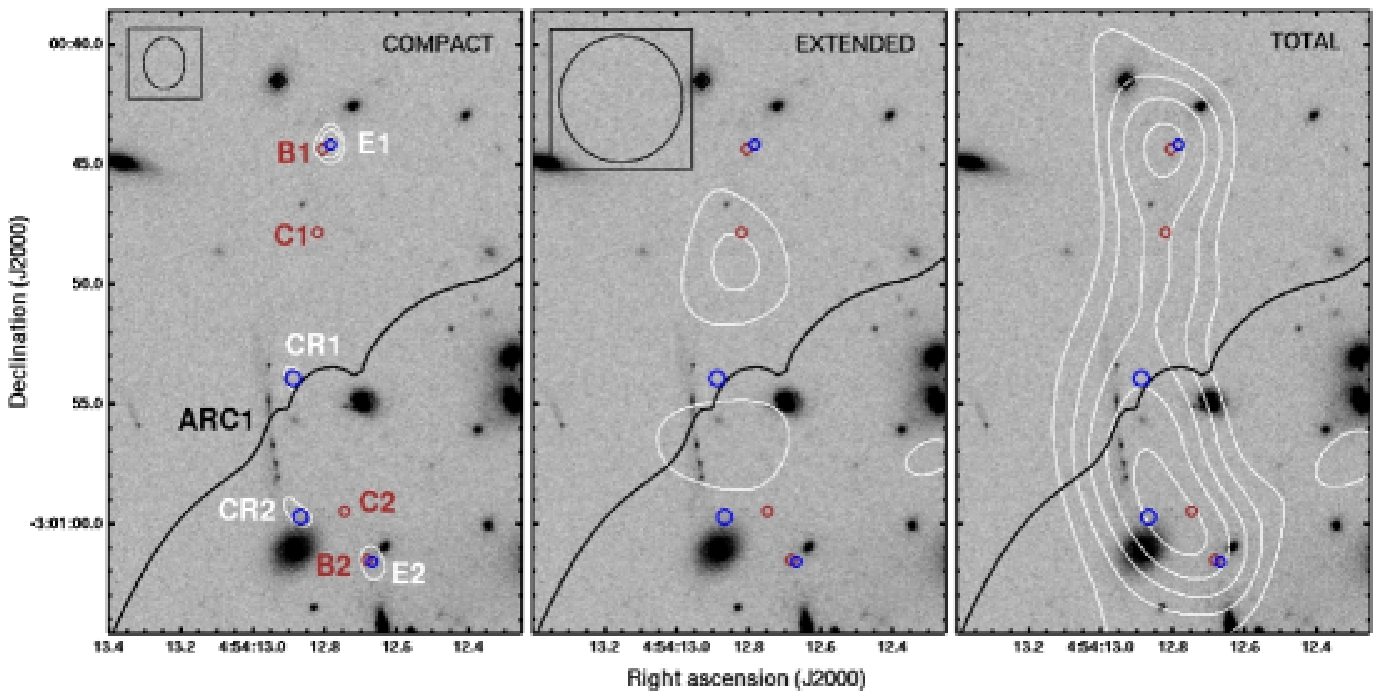}
  \caption{Detail of the HST image in the arc region, including the
  critical curve at $\rm z=2.911$ predicted by the lens model
  described in Sect. \ref{sec:modeling} (black line). The red circles
  indicate the positions of the multiply imaged EROs mentioned in
  \cite{BO04.1}, whereas the blue circles correspond to the compact
  radio detections indentified in this region. The radius of the
  circles indicate the estimated positional errors: $\sim 0.20\arcsec$
  for the EROs \citep[astrometric error for the standard stars quoted
  in][]{TA03.1}, between $0.1\arcsec - 0.2\arcsec$ for the radio
  sources (see Table \ref{tab_maxfit_R0.6}) and $\sim 0.08\arcsec$
  (half a pixel) for the optical sources. The contours correspond to
  the $\sim 2\arcsec$ resolution radio map presented in
  fig.\ref{A+Barray_R0.6} (left panel), and two of the $\sim 5\arcsec$
  resolution radio maps presented Fig. \ref{tap30_comparison} (central
  and right panel).}
  \label{merger}
\end{figure*}

\begin{figure}
  \centering
  \includegraphics[width=9cm,angle=0]{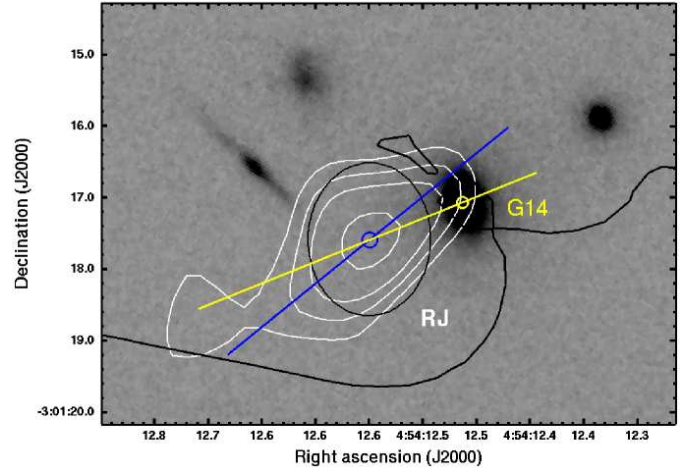}
  \caption{Detail of the HST image in the region of RJ, including the
  critical curve at $\rm z=2.911$ predicted by the lens model describe
  in sect.\ref{radio_vs_optical} (black line). The contours correspond
  to the $\sim 2\arcsec$ resolution radio map presented in
  fig.\ref{A+Barray_R0.6}. The radius of the circles show the
  estimated positional uncertainties: $0.1\arcsec$ for RJ (see Table
  \ref{tab_maxfit_R0.6}) and $\sim 0.08\arcsec$ (half a pixel) for
  G14. The blue line indicates the approximate orientation of the major
  axis of the radio source. For comparison, the yellow line shows the
  orientation of the RJ centroid respect to the galaxy G14. The black
  ellipse indicates the beam size of the radio map ($2.14\arcsec
  \times 1.71\arcsec$ at a position angle of $-1^{\circ}.43$).}
  \label{RJ}
\end{figure}

Figures \ref{merger}, \ref{RJ} and \ref{B3C3} show different details
of the HST image where the radio detections are located.  The
positions of the relevant sources are indicated with circles: blue for
the radio sources and red for the K' band sources. The counterparts of
the sub-mm emission proposed by B04 have been indicated following the
nomenclature introduced in Fig. \ref{intro}. Note that the images B3
and C3 are supposed to be unresolved in the K' band image presented in
B04, and therefore referred to as B3/C3.

The relative position of the radio and NIR sources in these figures
suggests that E1, E2 and E3 might be associated with B1, B2 and
B3/C3. To better quantify these associations, as well as identify
other possible counterparts, we have used Bayesian inference to
calculate the probability that the compact radio sources are
physically associated with their nearest optical/NIR sources. The
mathematical expression of this probability has been derived as
follows.

Lets $P(\mu|d)$ be the probability density of the true distance
\emph{$\mu$} between two sources given an observed distance
\emph{d}. Following Bayesian inference, this probability is given by:
\begin{equation}
P(\mu|d)= \frac{L(d|\mu) ~P(\mu)}{P(d)} 
\end{equation}

where $L(d|\mu)$ is the likelihood of the observed distance given a
true distance, $P(\mu)$ is the probability density of the true
distance before any measurement, and $P(d)$ is the probability density
of the observed distance for all possible distances (a normalization
constant).

As pointed out in \cite{CH06.1}, the likelihood of the observed
distance between two sources which positions are described by a
Gaussian distribution is not Gaussian. Instead, this likelihood in two
dimensions is given by:
\begin{equation}
  L(d|\mu)=\frac{d~I_{0}~(d~\mu/(\sigma_{1}^{2}+\sigma_{2}^{2}))}{\sigma_{1}^{2}+\sigma_{2}^{2}}~exp\left(\frac{-\mu^{2}-d^{2}}{2~(\sigma_{1}^{2}+\sigma_{2}^{2})} \right)
\end{equation}

Where $\sigma_{1}$ and $\sigma_{2}$ are the errors in the position of
each source, and $I_{0}$ is the modified Bessel function of integer
order zero.

Since the area inside an annulus at radius $\mu$ increases as $\mu$
d$\mu$, we assume that, for a source located at a random position,
$P(\mu)=\mu$. Therefore:
\begin{equation}
P(\mu|d) \propto  L(d|\mu) \cdot \mu 
\end{equation}

Which normalized gives:
\begin{equation}
B(\mu|d)= \frac{ L(d|\mu) \cdot \mu}{\int_0^\infty L(d|\mu) ~ d\mu}
\end{equation}

The integral of this function between $\mu_{1}$ and $\mu_{2}$ gives
the probability that, for a given pair of source centroids with
observed offset $d$, the true distance between them lies between
$\mu_{1}$ and $\mu_{2}$. Intuitively, one would say that the condition
for two sources to be counterparts of each other is $\mu=0$. The
problem is that, in the case of extended sources like galaxies, the
peak of emission at different wavelenghts can be located at different
positions, which means that $\mu \neq 0$ despite the fact that they
are physical counterparts. For this reason, we will consider that a
radio source has an optical/NIR counterpart when the centroid of the
source emission lies inside the area of the optical/NIR source. The
probability of meeting this condition is given by:
\begin{equation}
P_{assoc.}=\int_0^{r_{src}} B(\mu|d)~d\mu
\end{equation}

\begin{table}
\centering

\caption{ \textbf{Conunterparts of the radio detections.} The columns
show: possible counterparts of the radio detections (Optical/IR
source), mesured offset between the radio source and the optical/IR
source ($d_{\rm obs}$), radius of the optical/IR source ($r_{\rm
src}$) and probability that the centroid of the radio source lies
inside the area of the optical/IR source ($P_{\rm assoc.}$). The
coordinates of the counterpart sources can be found in tables
\ref{tab_catalog} and \ref{tab_constrains}. Sources which positions
were taken from the K' band image are indicated with an asterisk
(otherwise we used the HST coordinates). The positions of the radio
sources used to calculate $d_{\rm obs}$ correspond to the values
derived with {\scriptsize MAXFIT} in the aligned radio map.}

\begin{tabular}{lcccccc}
\hline\hline
\noalign{\smallskip}
Radio  & Optical/IR  & $d_{\rm obs}$  & $r_{\rm src}$  & $P_{\rm assoc.}$\\ 
source & source      & (arcsecs)  & (arcsecs)  & ($\%$)          \\
\noalign{\smallskip}
\hline 		
\noalign{\smallskip}        
E1     & B1$^\ast$   & 0.38       &  0.6       & 67.8          \\
\noalign{\smallskip}
\hline 		
\noalign{\smallskip}	       	
CR1    & ARC 1.1     & 1.12       &  0.15      & $< 1$        \\
       & ARC 2.1     & 1.40       &  0.25      & $< 1$       \\  
       & ARC 3.1     & 1.60       &  0.20      & $< 1$        \\
\noalign{\smallskip}
\hline 		
\noalign{\smallskip}		          
CR2    & G2          & 1.39       &  0.70       &  $< 1$         \\ 
       & C2$^\ast$   & 1.82       &  0.60       &  $< 1$       \\
       & ARC 1.2     & 1.92       &  0.15       &  $< 1$     \\
\noalign{\smallskip}
\hline 		
\noalign{\smallskip}		          
E2     & B2$^\ast$   & 0.26       &  0.60       & 79.29        \\
       & G10         & 0.74       &  0.23       & $< 1$     \\
\noalign{\smallskip}
\hline 		
\noalign{\smallskip}		            
E3     & 4.3$^\ast$   & 0.77       &  0.60       & 20.7         \\
       & 5.3$^\ast$   & 1.22       &  0.30       & $< 1$        \\ 
       & G23         & 1.54       &  0.35       & $< 1$      \\ 
\noalign{\smallskip}
\hline 	
\noalign{\smallskip}		            
RJ     & G8          & 1.91       &  0.30       & $< 1$    \\
       & G11         & 2.40       &  0.26       & $< 1$    \\
       & G14         & 1.39       &  0.60       & $< 1$    \\			   
\hline

\end{tabular}
\label{offsets}
\end{table}

A list with the nearest optical/NIR sources of each radio detection,
together with their associated probabilities, is presented in Table
\ref{offsets}. As expected, only the images of ERO B have a
non-negligible probability of being associated with some of the radio
sources (E1, E2 and E3). However, if E1, E2 and E3 are indeed the
radio counterparts of B1, B2 and B3, they have to be multiple images
produced by a radio source at $z\sim2.9$. A quantitative analysis of
this lensing scenario is presented in Sect. \ref{lensmodel_radio}.

Finally, note that no optical/NIR counterpart has been identified with
RJ. Therefore, given the depths of the optical and NIR images, it
seems more likely that the extended morphology of RJ corresponds to an
AGN rather than a resolved low redshift star forming galaxy. Within
the AGN scenario, we expect that the peak of the radio emission
corresponds to the position of an undetected optical source, and the
extensions in opposite directions are two jets coming from it. Another
conceivable scenario would be that the galaxy G14 is an AGN host with
a one sided radio extension.  However, the mayor axes of RJ does not
seem to be aligned with G14, and this kind of AGN is not very common.

\subsection{Comparison between radio and sub-mm}

\begin{figure*}
  \centering
  \includegraphics[width=18cm,angle=0]{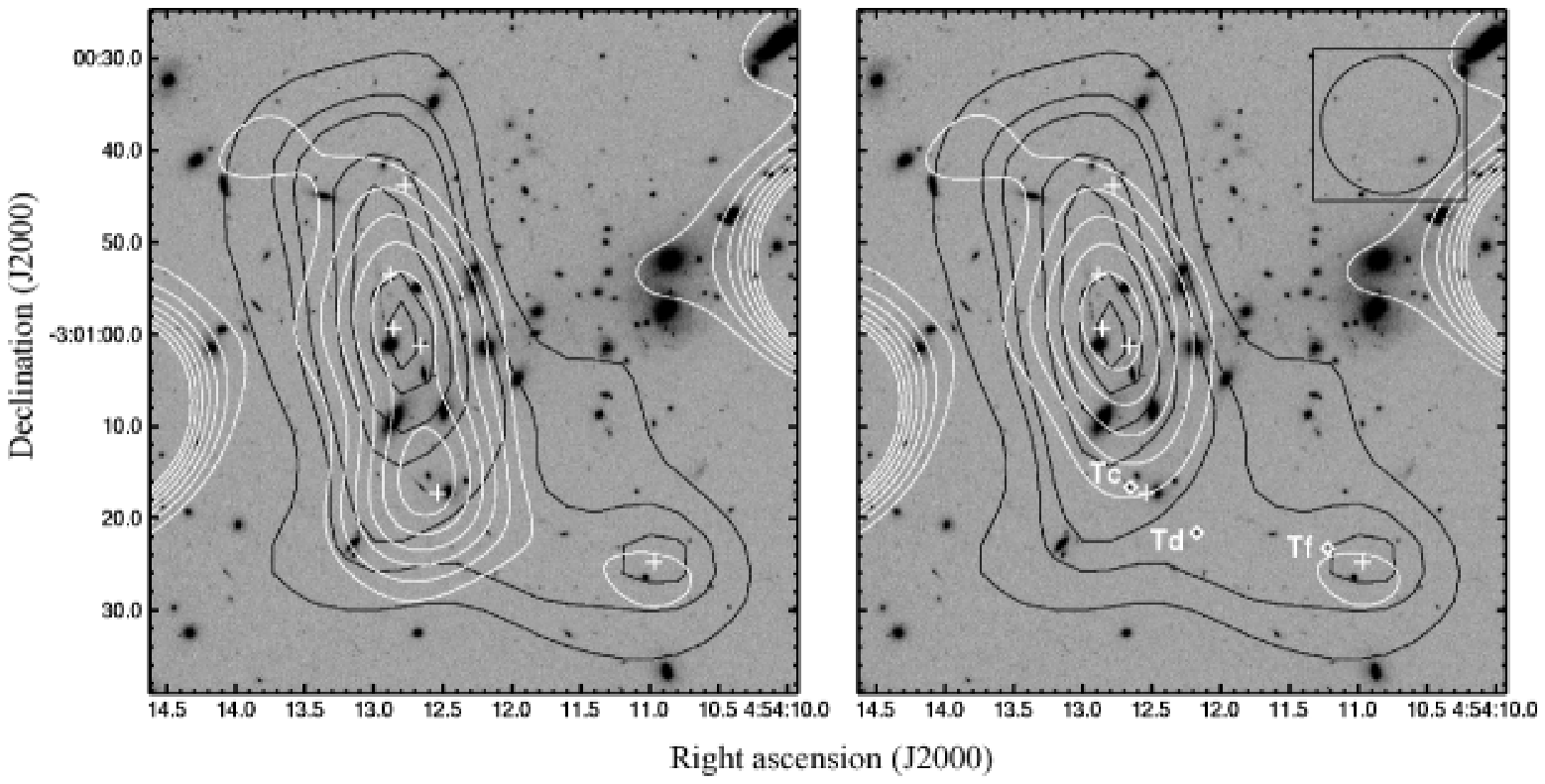}
  \caption{ Detail of the HST image of the cluster center, with the
  sub-mm 850~$\mu$m contour map superimposed in black. The white lines
  correspond to the VLA 1.4~GHz contour map obtained after tapering
  the A+B-array data to match the resolution in sub-mm. Radio contours
  are drawn at 2, 3, 4, 5, 6 \& 7 times the $1\sigma$ noise level of
  33.93 $\mu$Jy~$\rm beam^{-1}$. Sub-mm contours are drawn at 4, 6, 7,
  9, 10, 11 \& 11.5 mJy~beam$^{-1}$. The white crosses indicate the
  position of the radio sources identified within the sub-mm emission
  region. The beam size ($15\arcsec \times 15\arcsec$) is indicated in
  the top right corner. \textbf{Left panel:} map produced including
  all radio detections. \textbf{Right panel:} map produced after
  subtracting the CC model of RJ from the data. Indicated with
  diamonds are the positions of 3 EROs reported in \cite{TA03.1}
  that could be contributing to the observed sub-mm emission.}
  \label{radio_vs_submm}
\end{figure*}

\begin{figure}
  \centering
  \includegraphics[width=8cm,angle=0]{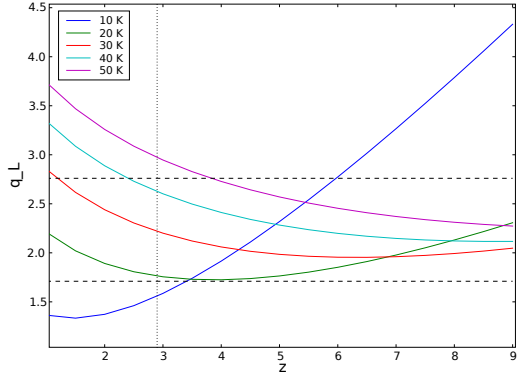}
  \caption{Change of the parameter $q_{\rm L}$ with redshift for
  different characteristic dust temperatures. The horizontal dashed
  lines indicate the minimum and maximum value of $q_{\rm L}$ found in
  the SMG sample studied by \cite{KO06.1}. The dotted vertical line
  indicates the position of z=2.9.}
  \label{ql_vs_z}
\end{figure}

In this section, we will use the FIR-radio correlation to check
whether the radio detections could be associated with the observed
sub-mm emission. To make this kind of analysis, it is necessary to
match the resolutions of the radio and sub-mm observations. Therefore,
the A+B array data was tapered with a Gaussian of
$8.5~K\lambda$\footnote{which corresponds to a clean beam of
$15.78\arcsec \times 14.65\arcsec$} and restored with a clean beam of
$15\arcsec \times 15\arcsec$ at the end of the imaging process. The
left panel of Fig. \ref{radio_vs_submm} shows the final tapered radio
map (white contours) superimposed upon the sub-mm map (black contours)
and the HST image of the cluster core. The positions of the detections
in the high resolution radio map have been indicated with crosses as
reference.

If all the observed sub-mm emission would be produced by a single SMG
that does not host an AGN, it is expected that the radio and sub-mm
morphologies would resemble each other. This is because the origin of
the FIR-radio correlation seems to be linked to massive star
formation\footnote{while young massive stars produce UV radiation that
is re-emitted in the FIR by the surrounding dust, old massive stars
die as SN producing electrons that are accelerated by the galactic
magnetic field generating synchrotron radio emission.}, which means
that both the radio and sub-mm emission are originated in
(approximately) the same regions of the galaxy. However, the $S_{\rm
850~\mu m}/S_{\rm 1.4~GHz}$ flux ratio observed in sub-mm galaxies
displays a broad scatter which strongly depends on the characteristic
dust temperature and the redshift \citep[e.g][]{BL02.1,
CH05.1}. Therefore, if the observed sub-mm emission is being produced
by several blended SMGs, we might find ``morphological
inconsistencies'' like the one present in the left panel of
Fig. \ref{radio_vs_submm} (note that the brightest peak of the radio
emission is located in the region of RJ instead of been coincident
with the brightest sub-mm peak).

Interestingly, if the CC model of RJ is subtracted from the data, the
morphology of the radio emission becomes remarkably similar to the
sub-mm emission (right panel of Fig. \ref{radio_vs_submm}). This
result strongly suggests that the source/sources responsible for the
radio emission observed in this panel are also responsible for the
bulk of the sub-mm emission. The other conclusion derived from this
comparison is that, if RJ is contributing to the sub-mm emission,
either its properties (redshift and/or dust temperature) are different
from the properties of the source/sources associated with the other
radio detections, or it has a ``radio excess'' due to an AGN.

\begin{figure*}
  \centering
  \includegraphics[width=15cm,angle=0]{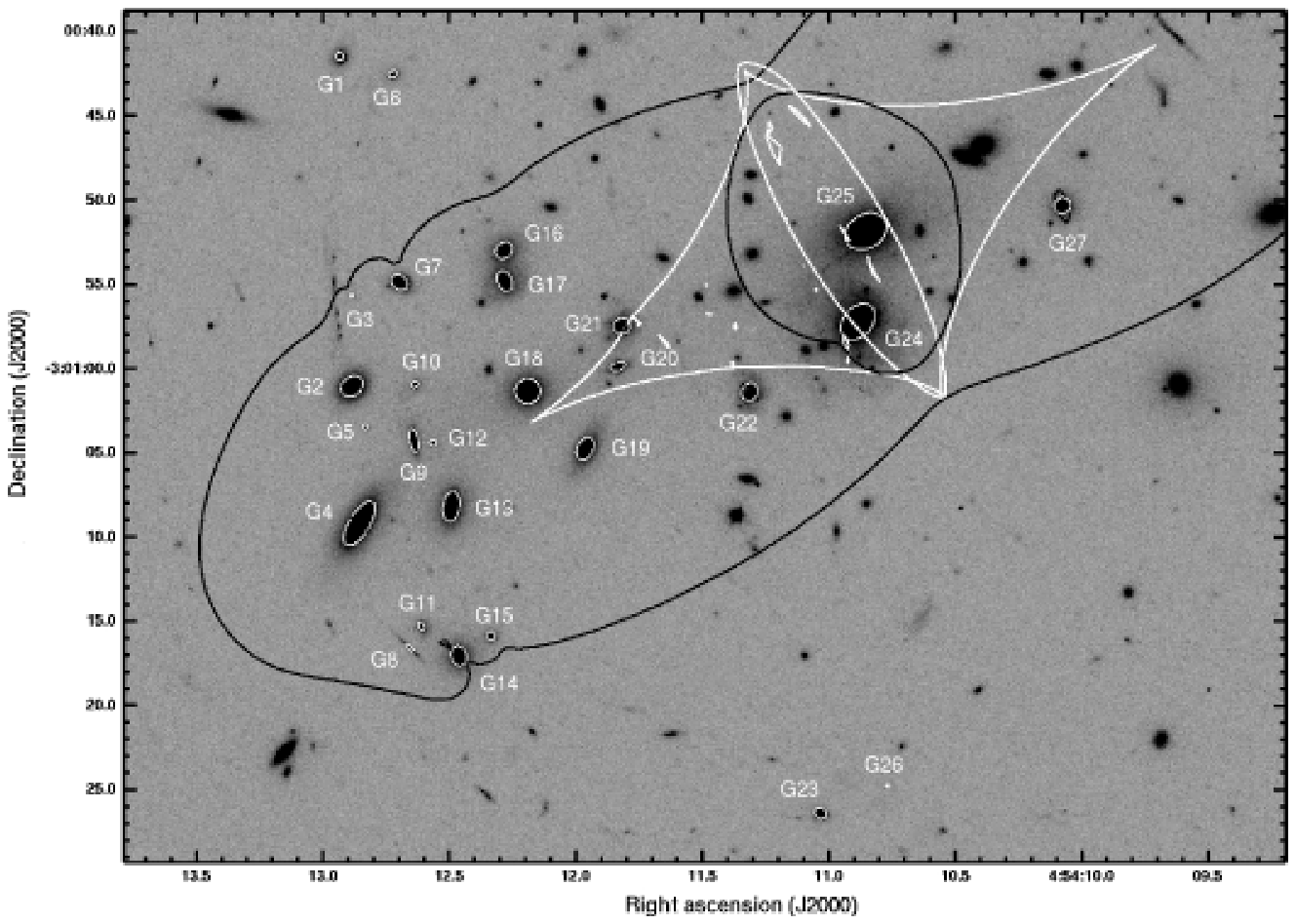}
  \caption{Detail of the HST image of the cluster core, indicating all
  the galaxies that have been included in the lens model. The sizes of
  the ellipses correspond to the morphological parameters listed in
  Table \ref{tab_catalog}. The lines represent the critical curves
  (black) and caustics (white) predicted by the lens model at $\rm
  z=2.911$.}
  \label{lensmodels}
\end{figure*}

Using $\rm 350~\mu m$, $\rm 850~\mu m$ and 1.4~GHz observations of 15
bright SMGs with spectroscopic redshifts, \cite{KO06.1} (K06
hereafter) conclude that the FIR-radio correlation remains valid for
SMGs with $z \sim 1-3$ and luminosities between $\rm
10^{11}-10^{13}~M_{\odot}$ (except when they are radio-loud AGN).  To
see how our system compares with this sample, we have made a more
quantitative analysis of the FIR-radio correlation using the $q_{\rm
L}$ parameter introduced in K06:

\begin{equation}
q_{\rm L}= log \left(\frac{ L_{\rm FIR}}{ [\rm 4.52~THz]~ L_{\rm
1.4GHz}} \right)
\end{equation}
 
were total FIR luminosity is inferred from the flux density at
$\rm 850~\mu m$ assuming a grey body model: 

\begin{equation}
L_{\rm FIR}= 4 \pi D_{\rm L}^2 \Gamma (4+\beta) \zeta (4+\beta) \left(
\frac{kT_{\rm d}}{h \nu} \right)^{4+\beta} \left( e^{h \nu/ kT_{\rm
d}} -1 \right) \nu S_{\rm 850\mu m}
\end{equation}

and the radio luminosity can be derived as:

\begin{equation} 
L_{\rm 1.4GHz}= 4 \pi D_{\rm L}^2 ~ S_{\rm 1.4GHz}~(1+z)^{\alpha -1}
\end{equation}

In these formulas, $D_{\rm L}$ is the luminosity distance, $\beta$ is
the effective emissivity index (we assume $\beta= 1.5$), $T_{\rm d}$ is
the characteristic dust temperature, $\nu$ corresponds to the observed
sub-mm frequency, and $\alpha$ is the radio spectral index (we assume
$\alpha=0.7$).

At the position of the sub-mm peak, the radio and sub-mm fluxes were
derived using the area delimited by the $3\sigma$ radio contour shown
in the left panel of Fig. \ref{radio_vs_submm} ($ S_{\rm 850\mu
m}=15.5\pm 2.8$~mJy and $S_{\rm 1.4GHz}=241 \pm 42$~$\mu$Jy). The
different values of $q_{\rm L}$ derived from these fluxes as function
of redshift and $T_{\rm d}$ are shown in Fig. \ref{ql_vs_z}. The
horizontal dashed lines indicate the minimum and maximum value of
$q_{\rm L}$ found in the K06 sample (discarding AGN hosts). Note that
there is a wide range of $z$ and $T_{\rm d}$ for which the observed
fluxes are consistent with K06. If we now assume that the radio and
sub-mm emission are produced by a (lensed) star forming galaxy at $\rm
z=2.9$, only temperatures between $\sim 20-40$~K would be allowed,
which is the temperature range in which most of the K06 sources lie.

Since AGNs do not follow the FIR-radio correlation, an estimate of the
$q_{\rm L}$ value for RJ could provide extra evidence to
confirm/discard the possible AGN nature of this source.  However, that
would require a higher resolution $\rm 850~\mu m$ map, to confirm the
connection of RJ with a discrete sub-mm source and get an accurate
estimate of its sub-mm flux. In addition, the redshift of RJ needs to
be determined in order to break the degeneracy between $z$ and $\rm
T_{d}$ illustrated in Fig. \ref{ql_vs_z}.

Finally, is important to mention that, although the energy output of
the majority of SMGs is dominated by star formation, about $30-50 \%$
host an AGN \citep{AL05.1}. Therefore, even if the source
associated with RJ is an AGN, it could still be contributing to the
sub-mm emission observed in that region. Other potential contributors
to the observed sub-mm emission are indicated in the right panel of
Fig. \ref{radio_vs_submm}.

\section{Gravitational lens modeling of MS0451} 
\label{sec:modeling}

\begin{table*}
\centering
\caption{ \textbf{Galaxy catalog for the new lens model of MS0451.}
The colums show: galaxy number (ID) galaxy coordinates (RA, DEC),
semi-major axes (a), semi-minor axes (b), position angle ($\theta$)
and magnitude in the HST image (mag). An asterisk in the ID number
indicate galaxies that were not included in the model presented in
\cite{BO04.1}.}

\begin{tabular}{lclcccc}
\hline\hline
\noalign{\smallskip}
ID & RA ($+4^{h}$ $54^{m}$)   & \multicolumn{1}{c}{DEC ($- 3^{\circ}$)}         & a        & b        & $\theta$ & mag    \\
   & J2000 (sec)              & \multicolumn{1}{c}{J2000 ($\arcmin$,$\arcsec$)} & (arcsec) & (arcsec) & (deg)    & (F814W) \\
\noalign{\smallskip}
\hline 
\noalign{\smallskip}
01$^{\ast}$ & 12.933 & $00:41.494$ & 0.24 & 0.23  &  65   & 21.9511 \\
02        & 12.884 & $01:01.056$ & 0.70 & 0.55  &  40   & 20.1111 \\
03$^{\ast}$ & 12.886 & $00:55.685$ & 0.12 & 0.09  & -28   & 24.6286 \\
04        & 12.855 & $01:09.205$ & 1.50 & 0.60  &  59   & 19.0257 \\
05$^{\ast}$ & 12.831 & $01:03.476$ & 0.17 & 0.15  &  55   & 23.2229 \\
06$^{\ast}$ & 12.723 & $00:42.529$ & 0.24 & 0.17  &  30   & 22.3768 \\
07        & 12.697 & $00:54.889$ & 0.53 & 0.40  &  140  & 20.8427 \\
08$^{\ast}$ & 12.655 & $01:16.586$ & 0.30 & 0.10  & -44   & 22.5806 \\
09$^{\ast}$ & 12.639 & $01:04.288$ & 0.70 & 0.25  & -80   & 20.7576 \\
10$^{\ast}$ & 12.634 & $01:00.949$ & 0.23 & 0.17  &  50   & 22.4744 \\
11$^{\ast}$ & 12.610 & $01:15.312$ & 0.26 & 0.16  & -69   & 22.1831 \\
12$^{\ast}$ & 12.563 & $01:04.426$ & 0.19 & 0.19  &  20   & 22.5221 \\
13      & 12.492 & $01:08.202$ & 0.90 & 0.50  &  81   & 19.6345 \\
14      & 12.461 & $01:17.075$ & 0.60 & 0.45  &  105  & 20.0909 \\
15$^{\ast}$ & 12.334 & $01:15.892$ & 0.25 & 0.25  & -15   & 21.9945 \\
16      & 12.283 & $00:53.001$ & 0.55 & 0.45  &  50   & 20.3429 \\
17      & 12.280 & $00:54.816$ & 0.64 & 0.42  &  115  & 20.0471 \\
18      & 12.189 & $01:01.348$ & 0.75 & 0.73  &  10   & 19.5202 \\
19      & 11.962 & $01:04.756$ & 0.70 & 0.46  &  63   & 20.1070 \\
20$^{\ast}$ & 11.833 & $00:59.879$ & 0.45 & 0.25  &  16   & 21.1736 \\
21      & 11.817 & $00:57.484$ & 0.53 & 0.45  &  15   & 20.3933 \\
22      & 11.312 & $01:01.400$ & 0.55 & 0.45  &  66   & 20.6069 \\
23$^{\ast}$ & 11.032 & $01:26.400$ & 0.35 & 0.30  & -40   & 21.4960 \\
24      & 10.884 & $00:57.231$ & 1.20 & 0.85  &  50   & 18.1360 \\
25      & 10.853 & $00:51.865$ & 1.30 & 1.00  &  30   & 18.1454 \\   
26$^{\ast}$ & 10.767 & $01:24.772$ & 0.12 & 0.08  &  21   & 25.2539 \\
27      & 10.076 & $00:50.338$ & 0.47 & 0.47  &  56   & 20.5351 \\
\hline 

\end{tabular}
\label{tab_catalog}
\end{table*}

In order to investigate the possible lensed nature of the radio
detections, we used the publicly available
{\scriptsize{\textsc{LENSTOOL}}}\footnote{http://www.oamp.fr/cosmology/lenstool/}
code to model the mass distribution of the cluster. This
modeling involves an optimization procedure, aimed to find the
mass model parameter values that best reproduce the observational
constrains (positions of the multiply imaged systems). For a detailed
description of the strong lensing methodology followed in the
{\scriptsize LENSTOOL} software we refer to \cite{LI07.1},
\cite{JU07.1} and the appendix A2 of \cite{SM05.1}.

The current best lens model of the cluster MS0451, published in B04
(B04 model hereafter), was produced with the first version of
{\scriptsize LENSTOOL} \citep{KN93.1}. This version is based on a
downhill $\chi^{2}$ minimization (which can be very sensitive to local
minimum in the likelihood distribution), and does not provide
estimates of the errors on the optimized parameters. The latest
{\scriptsize LENSTOOL} version \citep{JU07.1} used in this paper
includes a Bayesian Monte Carlo Markov chain optimization routine,
which allows to determine errors in the optimized parameters and
lowers the probability of ending in local $\chi^{2}$ minimum.

\subsection{Mass Model}

The cluster mass model is constructed using two different components:
(i) a cluster halo, which represents the DM component on cluster
scales and the baryonic intracluster gas, and (ii) 27 galaxy halos
that account for individual galaxies. Both components are described
using a PIEMD profile \citep[e.g.][]{LI05.1, EL07.1},
parametrized by position (RA, DEC), position angle ($\theta$),
ellipticity ($\epsilon$), core radius ($R_{\rm core}$), scale radius
($R_{\rm cut}$) and velocity dispersion ($\sigma_{0}$). The density
distribution for this profile is given by :

\begin{equation}
\rho(r)=\frac{\rho_{0}}{(1+r^2/R_{\rm core}^2)(1+r^2/R_{\rm cut}^2)}
\end{equation}

The galaxy-scale component contains the cluster members located
relatively close to the area were the multiple images are formed,
because they are the ones that have the strongest effect in the
potential of that region. We also included some fainter galaxies
located close to the strong lensing constraints, since they are known
to perturb the strong lensing configuration \citep{ME07.1}. The
details of our galaxy catalog are listed in Table \ref{tab_catalog},
and their positions and sizes are displayed in
Fig. \ref{lensmodels}. Note that the galaxy catalog used in the B04
model (39 galaxies in total) includes cluster members located farther
away, but not the faint galaxies mentioned before.

\begin{figure}
  \centering
  \includegraphics[width=7cm,angle=0]{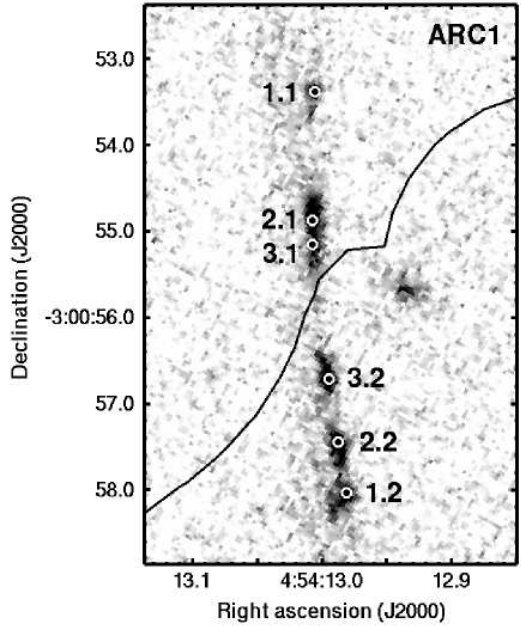}
  \caption{Detail of the optical arc as seeing in the HST image. The
  circles indicate the positions of the three sets of mirror images
  listed in Table \ref{tab_constrains}.}
  
\label{arc1}
\end{figure}

\subsection{Multiple images}

\begin{figure}
  \centering
  \includegraphics[width=8cm,angle=0]{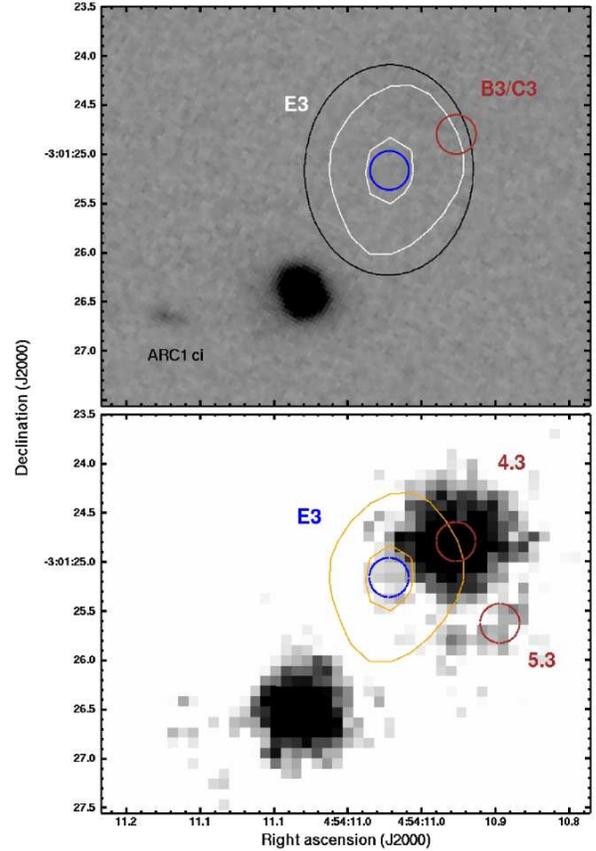}
  \caption{Detail of the source E3, including its radio contours as
  presented in fig.\ref{A+Barray_R0.6}.  \textbf{Top panel:} HST image
  of the region, including the positions of the counter images of the
  optical arc (ARC1 ci) and the ERO pair (B3/C3) reported by
  \cite{BO04.1}. \textbf{Bottom panel:} K' band image of the region,
  indicating the constraints used in the lens model (4.3 and 5.3). The
  radius of the circles show the estimated positional uncertainties:
  $\sim 0.20\arcsec$ for the K' band sources \citep[astrometric error
  for the standard stars quoted in][]{TA03.1}, $0.2\arcsec$ for E3
  (see Table \ref{tab_maxfit_R0.6}) and $\sim 0.08\arcsec$ (half a
  pixel) for the optical sources. The black ellipse indicates the beam
  size of the radio map ($2.14\arcsec \times 1.71\arcsec$ at a
  position angle of $-1^{\circ}.43$).}
  \label{B3C3}
\end{figure}

Figure \ref{arc1} shows a close up of ARC1 as seeing on the ACS
image. From its knotted structure we can identify 3 sets of mirror
images, each of which should have its own counter-image. However,
since it is not possible to distinguish the different knots in the
de-magnified image of the arc (ARC1 ci, see Fig. \ref{B3C3}), we
assigned the same position to the counter-image of each set of
constraints coming from the arc.

The positions for the EROs used in our model come from the Subaru K'
band image published in \cite{TA03.1}, which is deeper and has higher
resolution than the CFHT K' band image published by B04. Based on
their model, B04 identified the NIR source labeled B3/C3 as the result
of the blending between the counter-images of B and C. However, we
have detected a faint source (labeled `5.3' in Fig. \ref{B3C3})
located very close to B3/C3 in the Subaru K' band image. This source
was not reported in \cite{TA03.1} because it is too faint to be
classified as a DRG\footnote{Distant Red Galaxies, photometrically
defined as $J - Ks > 2.3$} (Tadafumi Takata, private
communication). Since it is possible that this source is the resolved
counter-image of ERO C, we decided to include it as a constraint in
the new model. The full set of constraints used for the optimization
is listed in Table \ref{tab_constrains}.

\begin{table}
\centering
\caption{ \textbf{Constraints for the new lens model of MS0451.}
Sources used as constraints in the lens model presented by
\cite{BO04.1} are indicated with an asterisk in their ID number.}

\begin{tabular}{ccclc}
\hline\hline
\noalign{\smallskip}
Source name & ID & RA ($+4^{h}$ $54^{m}$) & DEC ($- 3^{\circ}$) & Band \\
            &    & J2000 (sec)            & J2000 ($\arcmin$,$\arcsec$) &  \\
\noalign{\smallskip}
\hline 
\noalign{\smallskip}

      & 1.1$^\ast$ & 12.956 & $0:53.379$ & F814W \\
ARC1  & 1.2$^\ast$ & 12.931 & $0:58.029$ &               \\
      & 1.3$^\ast$ & 11.124 & $1:26.641$ &  \\
\noalign{\smallskip}
\hline 		
\noalign{\smallskip}	   
      & 2.1        & 12.958 & $0:54.869$ & F814W \\
ARC1  & 2.2        & 12.938 & $0:57.446$ &               \\
      & 2.3        & 11.124 & $1:26.641$ &               \\
\noalign{\smallskip}
\hline 	    
\noalign{\smallskip}   
      & 3.1        & 12.958 & $0:55.147$ & F814W \\
ARC1  & 3.2        & 12.945 & $0:56.710$ &               \\
      & 3.3        & 11.124 & $1:26.641$ &               \\
\noalign{\smallskip}
\hline 	
\noalign{\smallskip}		      
B1    & 4.1$^\ast$ & 12.806 & $0:44.344$ & K' band \\
B2    & 4.2$^\ast$ & 12.684 & $1:01.488$ &             \\
B3    & 4.3$^\ast$ & 10.927 & $1:24.795$ &             \\
\noalign{\smallskip}
\hline 	
\noalign{\smallskip}		      
C1    & 5.1        & 12.822 & $0:47.834$ & K' band \\
C2    & 5.2        & 12.747 & $0:59.481$ &             \\
C3    & 5.3        & 10.897 & $1:25.627$ &             \\
\noalign{\smallskip}
\hline 	           	      
\noalign{\smallskip}
RC1   & 6.1        & 12.890 & $0:53.907$ & 1.4~GHz \\
RC2   & 6.2        & 12.868 & $0:59.696$ &              \\
\hline 

\end{tabular}
\label{tab_constrains}
\end{table}

\begin{table}
\centering
\caption{Ranges in which the free parameters are allowed to vary
during the lens model optimization.}
\begin{tabular}{cc}
\hline\hline
\noalign{\smallskip}
Parameter & Opt. range    \\
\noalign{\smallskip}
\hline
\noalign{\smallskip}
X             &  $\pm 6\arcsec$ from BCG center \\
Y             &  $\pm 6\arcsec$ from BCG center\\
$\epsilon$    &  0 - 0.7 \\
$\theta$      &  0 - 180 deg\\
$\sigma_{0}$  &  900 - 1800 km/s\\
$R_{\rm core}$    &  30 - 120 kpc\\
\noalign{\smallskip}
\hline
\noalign{\smallskip}
$\sigma_{0}^{\ast}$  & 120 - 140 km/s \\
$R_{\rm cut}^{\ast}$     & 30 - 180 kpc \\
\noalign{\smallskip}
\hline
\end{tabular}
\label{tab_opt_ranges}
\end{table}

\subsection{Optimization}

\begin{table*}
\caption{\textbf{Most likely mass model parameters.} The columns show:
position respect to the brightest cluster galaxy (X,Y), ellipticity of
the mass distribution ($\epsilon$, expressed as $a^2-b^2/a^2+b^2$,
were $a$ and $b$ are the semi-major and semi minor axes of the ellipse
that describes the light distribution of the clump), position angle
($\theta$), velocity dispersion ($\sigma_0$), core radius ($r_{\rm
core}$) and scale radius ($r_{\rm cut}$). Error bars correspond to
$1\sigma$ confidence level as inferred from the \textsc{mcmc}
optimization.  Values into brackets are not optimized.  When the
posterior distribution is not Gaussian, we report the mode and
asymmetric error bars. The coordinates of the brightest cluster galaxy
are $\rm RA=4^{h} 54^{m} 10.87^{s}$, $\rm DEC=-3^{\circ} 0\arcmin
54.00\arcsec$.}
\begin{center}
\begin{tabular}{cccccccc}
\hline
\hline
\noalign{\smallskip}
Clump & X            & Y           & $\epsilon$ & $\theta$ & $\sigma_0$ & $r_{\rm core}$ & $r_{\rm cut}$  \\
      & ($\arcsec$)  & ($\arcsec$) &            & (deg)    &            & (kpc)          & (kpc)          \\
\noalign{\smallskip}
\hline
\noalign{\smallskip}
Cluster & -0.5$^{+4.1}_{-1.2}$  & 2.7$\pm$1.5 & 0.69$^{+0.0}_{-0.1}$ & 31.3$\pm$1.1 & 1144.8$\pm$36 & 60.8$\pm$24 & [1\,500] \\
\noalign{\smallskip}
L$^*$ elliptical galaxy & -- & -- &-- & -- & 135.1$\pm$43.9 & [0] & 30.1$^{+43.6}_{-4.9}$ \\
\noalign{\smallskip}
\hline
\end{tabular}
\end{center}
\label{lensmodel_results}
\end{table*}

\begin{table*}
\caption{\textbf{Best mass model.} The columns show: position respect
to the brightest cluster galaxy (X,Y), ellipticity of the mass
distribution ($\epsilon$, expressed as $a^2-b^2/a^2+b^2$, were $a$ and
$b$ are the semi-major and semi minor axes of the ellipse that
describes the light distribution of the clump), position angle
($\theta$), velocity dispersion ($\sigma_0$), core radius ($r_{\rm
core}$) and scale radius ($r_{\rm cut}$). Values into brackets are not
optimized. The coordinates of the brightest cluster galaxy are $\rm
RA=4^{h} 54^{m} 10.87^{s}$, $\rm DEC=-3^{\circ} 0\arcmin
54.00\arcsec$.}
\begin{center}
\begin{tabular}{cccccccc}
\hline
\hline
\noalign{\smallskip}
Clump & X             & Y             & $\epsilon$ & $\theta$ & $\sigma_0$ & $r_{\rm core}$ & $r_{\rm cut}$  \\
      & ($\arcsec$)   & ($\arcsec$)   &            & (deg)    & (km/s)     & (kpc)          & (kpc)          \\
\noalign{\smallskip}
\hline
\noalign{\smallskip}
Cluster & $-0.771414$  & 2.256631 & 0.695267 & 31.146240  & 1117.322994 & 7.667450 & [1\,500] \\
\noalign{\smallskip}
L$^*$ elliptical galaxy & -- & -- &-- & -- & 150.421411 & [0] & 3.880257 \\
\noalign{\smallskip}
\hline
\end{tabular}
\end{center}
\label{lensmodel_best}
\end{table*}

The set of free parameters used during the optimization are: (i) all
the parameters that characterize the cluster halo (except $R_{\rm
core}$) and (ii) the velocity dispersion ($\sigma_{0}^{\ast}$) and
scale radius ($ R_{\rm cut}^{\ast}$) of a galaxy at $\rm
z=z_{cluster}$ with a typical luminosity $L^{\ast}$ that corresponds
to an observed magnitude of $\rm m=18.8$.

The $\sigma_{0}$ and $R_{\rm cut}$ of all the galaxy halos included in
the mass model are derived from the luminosity of their associated
galaxy using the following empirical scaling relations:

\begin{equation}
     R_{\rm cut}=R_{\rm cut}^{\ast}    \left( \frac{L}{L^{\ast}} \right) ^{1/2}  \quad
  \sigma_{0}=\sigma_{0}^{\ast} \left( \frac{L}{L^{\ast}} \right) ^{1/4}
\end{equation}

The scaling relation for $\sigma_{0}$ assumes that mass traces light,
and its origin resides in the Tully-Fisher and Faber-Jackson
relations. The scaling relation for the radial parameter assumes that
the mass-to-light ratio is constant for all galaxies. The zero point
of these relations is set by the magnitude of the $L^{\ast}$
galaxy. The rest of the parameters of all galaxy halos (RA, DEC,
$\theta$ and $\epsilon$) are fixed to the values measured from the
light distribution of the related galaxy, while their $R_{\rm core}$
is set to zero.

The ranges in which the free parameters are allowed to vary during the
optimization are listed in Table \ref{tab_opt_ranges}. Note that the
scale radius describes the properties of the mass distribution on
scales much larger than the radius over which the multiple images can
be found. Therefore, since strong lensing cannot give any reliable
constrains on this parameter, its value is set to an arbitrary large
number (1500 kpc). The ranges adopted for the $L^{*}$ galaxy are
motivated by galaxy-galaxy lensing studies in clusters
\citep{NA98.1,NA02.2,NA02.1,LI07.1}.

Since the redshift of the EROs has not been confirmed
spectroscopically, the modeling procedure was carried out in four
steps. In the first step, only the sets of images from ARC1 at $\rm
z_{spect}=2.911$ where used as constraints in the optimization. The
resultant best model was then re-optimized, including the constraints
provided by ERO B but leaving $\rm z_{B}$ as free parameter. As a
result, the predicted redshift of ERO B is $\rm z_{B}=2.93 \pm 0.13$.
In the third step, the constraints from ERO C were used in a new
re-optimization of the best model obtained in the previous step,
assuming $\rm z_{B}=2.9$ and leaving $z_{C}$ as free parameter. This
provides a $\rm z_{C}=2.85 \pm 0.06$, which is the same redshift that
B04 derived for the ERO pair. In the final step, the previous best
model is re-optimized (in the \emph{image} plane) including all the
constraints with $\rm z=2.911$.

The Bayesian MCMC optimization routine included in the new version of
{\scriptsize{\textsc{LENSTOOL}}} provides two kinds of outputs: (i)
the likelihood of reproducing the observed constraints, independently
derived for each free parameter of the mass model, and (ii) the set of
model parameters that provides the best fit to the input data. The
most likely values for the model parameters obtained from these
histograms are listed in Table \ref{lensmodel_results}, whereas the
parameters of the best model ({\scriptsize{\textsc{MFINAL}}}
hereafter) are listed in Table \ref{lensmodel_best}.

The image positions are well reproduced, with image plane positional
rms differences between $\sim 0.1\arcsec$ and $0.3\arcsec$ (see Table
\ref{tab_lensmodel_imgs}). The projected mass within the Einstein
radius (here approximated by the ARC1 distance from the center, is
$M_{\rm 2D}~(30\arcsec) = 1.73 10^14 M_{\odot}$.

\begin{table}
\centering
\caption{Results of the best models for each set of constraints. The
columns show: ID of each set of constraints (System), reduced $\chi^2$
obtained for each model ($\chi^2$ final, zfix and zfree) and average
separation between the observed and predicted position of the lensed
images for each model (rms final, zfit and zfree).}
\begin{tabular}{ccccccc}
\hline\hline
\noalign{\smallskip}

System & $\chi^2$ & $\chi^2$ & $\chi^2$ & rms   & rms  & rms     \\
       & {\scriptsize{\textsc{MFINAL}}}    & {\scriptsize{\textsc{MZFIX}}}     & {\scriptsize{\textsc{MZFREE}}}    & {\scriptsize{\textsc{MFINAL}}} & {\scriptsize{\textsc{MZFIX}}} & {\scriptsize{\textsc{MZFREE}}}         \\
\noalign{\smallskip}
\hline 
\noalign{\smallskip}
1 & 3.89 & 6.71  & 4.04 & 0.34 & 0.45 & 0.35 \\ 
2 & 2.02 & 0.68  & 1.28 & 0.25 & 0.14 &	0.20 \\
3 & 1.33 & 0.74  & 0.95 & 0.20 & 0.15 &	0.17 \\
4 & 3.86 & 8.53  & 6.32 & 0.34 & 0.51 &	0.44 \\
5 & 0.50 & 1.48  & 0.62 & 0.12 & 0.21 & 0.14 \\      
6 &	 & 35.46 & 3.70 &      & 1.26 &	0.41 \\
\hline 
\end{tabular}
\label{tab_lensmodel_imgs}
\end{table}

\subsection{Analysis of the radio data using the new lens model}
\label{lensmodel_radio}

\begin{figure*}
  \centering
  \includegraphics[width=18cm,angle=0]{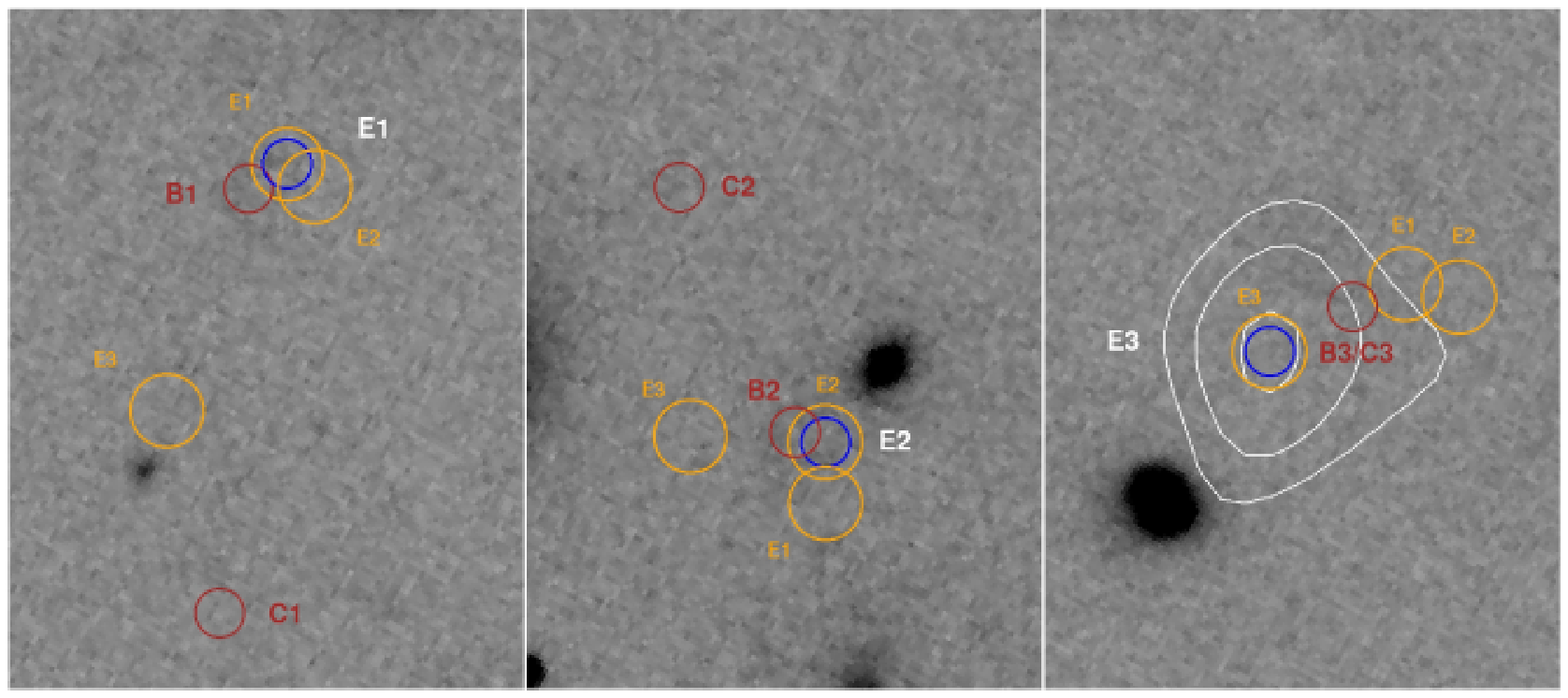}
  \caption{Analisis of the lensing nature of E1, E2 and E3 with the
  lens model {\scriptsize{\textsc{MFINAL}}}. The orange circles
  indicate the image positions predicted after tracing each of the
  three radio sources independently into the source plane and lens
  them back into the image plane. The radius of the circles correspond
  to $0.3\arcsec$. The rest of the labels are the same used in
  Fig. \ref{merger}.}
  \label{lensmodel_E3}
\end{figure*}

Since CR1 and CR2 are located in a region with several multiply-imaged
systems (see Fig. \ref{merger}), it is conceivable that they might
also constitute a set of mirror images. To test this hypothesis, we
redo the optimization using the positions of these two radio sources
as an additional set of constraints (system 6, see Table
\ref{tab_constrains}), first assuming $\rm z_{CR}=2.911$ and then
considering $\rm z_{CR}$ as a free parameter. A summary of the
properties of the resultant best models ({\scriptsize{\textsc{MZFIX}}}
and {\scriptsize{\textsc{ MZFREE}}} hereafter) is presented in Table
\ref{tab_lensmodel_imgs} and Table \ref{tab_lensmodel_radio}.

The optimization with z free results in a predicted redshift for
system 6 of $\rm z_{CR}=2.4 \pm 0.074$, which is not consistent with
$z=2.911$. If we compare the probability of
{\scriptsize{\textsc{MZFIX}}} and {\scriptsize{\textsc{ MZFREE}}} to
reproduce the observations (the evidence, see Table
\ref{tab_lensmodel_radio}), it is clear that {\scriptsize{\textsc{
MZFREE}}} is preferred over {\scriptsize{\textsc{MZFIX}}}. Note also
that the rms difference between the observed and predicted image
positions for system 6 is three times worse for
{\scriptsize{\textsc{MZFIX}}} compared with {\scriptsize{\textsc{
MZFREE}}} (see Table \ref{tab_lensmodel_imgs}). On the other hand, the
total $\chi^{2}$ of {\scriptsize{\textsc{ MZFREE}}} for all the
optical/IR multiply imaged systems ($\chi^{2}$ optical, see Table
\ref{tab_lensmodel_radio}) is just slightly worse that the
corresponding ~$\chi^{2}$ of {\scriptsize{\textsc{MFINAL}}}, which
means that {\scriptsize{\textsc{MFINAL}}} and {\scriptsize{\textsc{
MZFREE}}} are equally good. Note that the evidence cannot be used to
compare these two models because the number of free parameters in each
of them is different.

Therefore, unless the positional uncertainty of CR1 and CR2 could be
as large as $\sim 1\arcsec$ (which seems unlikely, since the
positional error for these sources derived from the observations is
$0.2\arcsec - 0.3\arcsec$), these results indicate that the lens model
favors an scenario in which CR1 and CR2 (if assumed to be mirror
images) are produced by a radio source that is not located at the same
redshift as the optical arc. However, as discussed in
Sect. \ref{sec:compact_vs_extended}, CR1 and CR2 seem to be two
relatively compact ($\sim 2\arcsec$) regions of an extended ($\sim
5\arcsec$) radio source. Therefore, it is possible that this extended
radio emission is a multiply imaged structure produced by a source at
$z=2.9$, in which CR1 and CR2 are not mirror images. Until the
structure of this extended radio emission can be robustly mapped with
deeper observations and included in the modeling process, the
posibility that CR1 and CR2 are associated with a lensed source at
redshift 2.9 remains open.

\begin{table}
\centering
\caption{General results of the best models. The columns show:
Nickname of the model (Model), sum of the reduced $\chi^2$ of all the
optical/NIR sets of constraints ($\chi^2$ optical), reduced $\chi^2$
for the radio set of constraints ($\chi^2$ radio), average separation
between the observed and predicted position of all the lensed images
used as constraints (aver rms), probability of the model to reproduce
the observations (log(evidence)).}
\begin{tabular}{ccccc}
\hline\hline
\noalign{\smallskip}
Model & $\chi^2$ optical & $\chi^2$ radio & aver rms & log(evidence) \\
\noalign{\smallskip}
\hline 
\noalign{\smallskip}
{\scriptsize{\textsc{MFINAL}}}  & 11.60 & $-$   & 0.26 & -24.503 \\
{\scriptsize{\textsc{MZFIX}}}   & 18.14 & 35.46 & 0.53 & -45.340 \\
{\scriptsize{\textsc{MZFREE}}}  & 13.21 & 3.70  & 0.30 & -29.741 \\
\hline
\end{tabular}
\label{tab_lensmodel_radio}
\end{table}

As mentioned in Sect. \ref{radio_vs_optical}, the multiple images of
ERO B have a non-negligible probability of being associated with the
radio detections E1, E2 and E3. However, the probability that E3 and
B3/C3 are associated is significantly lower than for E1-B1 and
E2-B2. For this reason, we used {\scriptsize{\textsc{MFINAL}}} to
check if the positions of E1, E2 and E3 would be consistent with a set
of three images produced by a radio source at $z=2.9$.  This was done
in the following way: for each radio detection, its position was traced
back into the source plane, and the resultant source was lensed again
into the image plane. In this way, the model provided three predicted
positions for each radio source, one of them being the position used
as input.

The results (shown in Fig. \ref{lensmodel_E3}) indicate that E3 cannot
be interpreted as the radio counterpart of B3/C3, since the images
produced by E1 and E2 in the counter image region are located on the
right side of B3/C3 (unlike in the case of E1 and E2), at a distance
which is inconsistent with the average rms of
{\scriptsize{\textsc{MFINAL}}}. A possible scenario that can explain
this result is that E1 and E2 are multiple images (but we don't detect
the associated counter-image because it is too faint) and E3 is a
non-lensed radio galaxy that it is serendipitously lying in this
region. However, as it will be discussed in the next section, it is
also possible to explain E3 as a lensed image produced by a source at
$z=2.9$ under the merger scenario proposed in B04.

\section{Discussion: the merger scenario}

Figure \ref{merger} shows a detail of the HST image at the position of
the sub-mm peak, where ARC1 and the EROs are located. The contour map
of the right panel indicates the total radio emission observed in this
region, whereas the left and central panels show the compact and
extended component. The black line correspond to the critical curve at
$z=2.9$ predicted by the lens model of the cluster described in
Sect. \ref{sec:modeling}.

If ARC1 and the EROs are indeed multiple images produced by three
different regions of a merger at $z=2.9$, the rest of the merger lying
between them should also be multiply-imaged. Hence, if the merger
contains a lot of dust (as expected from a high-z SMG), this images
would not be seen in the optical, but they would show up in the radio.

Under this scenario, the fact that CR1 and CR2 do not seem to have
optical counterparts, and lie in the region between ARC1 and the ERO
images, suggests that they could be lensed images associated with the
highly-obscured center of the merger. However, as discussed in
Sect. \ref{lensmodel_radio}, they can also be interpreted as a set of
mirror images produced by another radio galaxy located at a different
redshift. The real strong evidence in favor of the merger hypothesis
comes from the extended component in which CR1 and CR2 seem to be
embedded, which provides 38\% of the observed radio flux in the region
of the sub-mm peak. As is clearly shown in the central panel of
Fig. \ref{merger}, this extended component constitutes a bridge of
emission between the compact radio sources, as expected from the
multiply imaged dust-obscured material of the merger located between
the optical and NIR emitting regions. In complete agreement with this
scenario, the shape of the total radio emission in the right panel of
Fig. \ref{merger} follows the distribution of the optical arc and the
ERO images remarkably well.

In this case, we can also expect the extended emission to be multiply
imaged like ARC1 and the EROs, producing a counter-image in the region
between ARC1 ci and B3/C3 (see Fig. \ref{B3C3}). Since the images in
this region are de-magnified, it is likely that the expected radio
counterpart of B3/C3, and the counter image associated with the
extended emission, are blended. That would shift the peak of the total
observed radio emission towards ARC1 ci, explaining why the
probability derived for E3-B3/C3 is much lower than what it is found
for E1-B1 and E2-B2.

Finally, it is also not surprising that E1-B1 and E2-B2 are not
identified as robust ($\geq 95\%$) counterpart pairs. With the cluster
magnification generating a high resolution (distorted) view of the
inner structure of the merger, the observed offset between E1-B1 and
E2-B2 could be real, indicating that the NIR and radio emissions are
arising from slightly different regions inside the merger.

\section{Summary and Conclusions} 
\label{sec:conclusions}

SMM~J$04542-0301$ is an elongated region of bright sub-mm emission
located in the core of the cluster MS$0451.6-0305$. It has been
suggested \citep{BO04.1} to be partially produced by a multiply-imaged
$z=2.9$ merger which contains a LBG (lensed as ARC1 and ARC1 ci), and
two EROs (lensed as B1, B2, C1, C2 and B3/C3). Given the low
resolution and poor positional accuracy of the sub-mm map, it is very
difficult to confirm the connection between SMM~J$04542-0301$ and the
optical/NIR lensed images. However, high resolution radio
interferometric observations can help to establish this connection
thanks to the FIR-radio correlation, which has being found to be valid
for SMGs \citep{KO06.1, VL07.1, IB08.1, MI09.1}.\\

In a previous paper \citep{BE07.1}, we reported on the detection of
1.4~GHz radio emission coincident with this system using VLA archival
data. Now, following a more sophisticated data reduction procedure,
the previous B-array observations have being re-reduced and combined
with new high resolution A-array observations. The resultant data set
has been used to produced two radio maps of the cluster core:
\begin{itemize}
\item[(i)] a deep ($\sim10 ~\mu$Jy), high resolution ($\sim2\arcsec$)
map, to compare with the optical/NIR images. In this map we have
detected 5 sources (E1, E2, CR1, CR2 and E3) located near the
optical/NIR lensed images, and one source (RJ) located in the central
region of the sub-mm emission (see Fig. \ref{A+Barray_R0.6}). Despite
the fact that some of the sources have $\rm SNR < 5$, the fact that
the $-4\sigma$ peaks lie far from the central target region adds
confidence to the reality of the sources.\\

\item[(ii)] a $15\arcsec$ resolution map, to compare with
SMM~J$04542-0301$ (see Fig. \ref{radio_vs_submm}).\\
\end{itemize}

The source RJ is the brightest and most extended of these six
detections, with no obvious counterpart in the optical/NIR. Although
its extended morphology could be interpreted as an AGN signature, the
degeneracy between redshift and dust temperature allows the observed
$S_{\rm 850 \mu m}/S_{\rm 1.4~GHz}$ at the position of RJ to be
consistent with a starforming SMG that follows the FIR-radio
correlation. In any case, the low resolution of the $\rm 850~\mu m$
map makes impossible to determine if RJ is really contributing to the
observed sub-mm emission in this region.\\

On the other hand, the other five radio detections constitute a strong
observational evidence in favor of the merger scenario proposed by
\cite{BO04.1}. The evidence that supports this conclusion can be
summarized as follows:

\begin{itemize}

\item if the optical arc and the EROs are multiple images produced by
  three different regions of a merger at $z=2.9$, the dust obscured
  material between these regions is expected to be lensed in the same
  way, but only visible at radio and sub-mm. In agreement with this
  scenario, the shape of the radio emission observed at the sub-mm peak
  follows the distribution of the optical arc and the ERO images
  remarkably well (Fig. \ref{tap30_comparison}).\\

\item CR1 and CR2 do not have any optical/NIR counterpart. However,
  they seem to constitute two relatively compact emitting regions
  embedded in a $\sim 5\arcsec$ extended source located between E1 and
  E2. The presence of this extended component (which contributes 38\%
  of the total radio flux in the sub-mm peak) can only be explained if
  it is being produced by the dust obscured lensed material in the
  center of the merger.\\

\item The compact sources E1 and E2 have a high probability of being
  associated with the images of ERO B, but this probability is
  considerably lower for E3-B3/C3. A plausible explanation of this
  result can be provided by the merger scenario, in which the lensed
  extended component is expected to have a counter image which would
  be blended with the radio emission associated with E3/C3. This would
  shift the position of the total observed peak, lowering the
  probability of identifying E3 and B3/C3 as counterparts.\\

\item The fact that B1, B2 and B3/C3 are not robust ($\geq 95\%$)
  counterparts of E1, E2 and E3, indicates that the radio and NIR
  emission is being produced at slightly different positions in the
  sources plane, which can be distinguished thanks to the enhanced
  resolution provided by the lensing magnification.\\

\item If the $15\arcsec$ resolution radio map is produced after the CC
  model of RJ is subtracted from the data, its morphology turns out to
  be remarkably similar to the sub-mm map (right panel of
  Fig. \ref{radio_vs_submm}). This result strongly suggests that the
  sources E1, E2, CR1, CR2 and E3 are associated with the extended
  sub-mm emission, allowing us to establish a direct link between
  SMM~J$04542-0301$ and the merger.

\end{itemize}

In this paper we have also presented a new lens model of the cluster
MS$0451.6-0305$, produced with the latest version of the {\scriptsize
LENSTOOL} code \citep{JU07.1}, in which the positions of the ERO
images and the arc knots have been used as constrains. The redshifts
of the EROs predicted by this model are $z_{\rm B}=2.93 \pm 0.13$ and
$z_{\rm C}=2.85 \pm 0.06$, in agreement with \cite{BO04.1}.

Including the positions of CR1 and CR2 as mirror images in the model
significantly degrades the overall rms of the optimization. However,
since these two sources are embedded in an extended source (which
cannot be included in the lens model due to the current limitations of
the {\scriptsize LENSTOOL} code), it is very difficult to establish a
one-to-one identification of structures in the image plane. Therefore,
the possibility that CR1 and CR2 are lensed images produced by the
dust-obscured core of the merger remains open.

\section{Future prospects}

Radio observations of SMGs have been traditionally used to identify
their optical/NIR counterparts, but also to provide estimates of their
sizes \citep{CH04.1, BI08.1}. On the other hand, the magnification
bias due to the gravitational lensing effect produced by clusters of
galaxies has been used to increase the detection rate in sub-mm
surveys, and constrain the faint flux end of the submm counts
\citep[e.g.][]{KN08.1}. The observations presented in this paper,
however, illustrate that the prospects for both radio interferometry
and strong gravitational lensing in clusters of galaxies to study the
internal structure of sub-mm galaxies (SMGs) are also very
promising. With the improved sensitivity of e-MERLIN and EVLA, and the
revolutionary view of the sub-mm sky that will be provided by ALMA,
the enhanced resolution of multiply-imaged submm galaxies might permit
us, among other things, to resolve for the first time their different
star-forming regions, and help to assess the connection between
starburst and AGN processes that seem to coexist in these systems.

SMM$04542-0301$ constitutes a very promising system to carry out this
kind of studies, but there is still one fundamental piece of
information that is missing: redshift information for the NIR, radio
and submm components. Current millimeter facilities can provide
redshift information by detecting (for instance) CO rotational lines,
but the relatively narrow bandwidths of their receivers ($<$ 10 GHz
compared with the 115 GHz spacing between CO rotational lines) makes
the search of multiple lines very time consuming, and a prior estimate
of the redshift of the source is required. Fortunately, blind CO
redshift observations are now possible thanks to Z-Spec, a broadband
($\sim$115 GHz) millimeter/sub-millimeter spectrograph that has been
commissioned at the 10.4-m Caltech Submillimeter Observatory
\citep{BR09.1}. Once all the sources that are contributing to the
observed sub-mm emission of SMM$04542-0301$ have been robustly
identified, the increased bandwidth and sensitivity of the new
receivers that have been installed on the IRAM\footnote{Institut de
Radioastronomie Millimetrique} 30 m telescope (located on the Pico
Veleta, Spain) and the Plateau de Bure Interferometer (located in the
French Alps), will be well placed to characterize the physical
properties of this system in detail.

Finally note that, since these kind of systems consist of multiple
distorted images of the lensed object, a reliable reconstruction of
the source is essential for their interpretation. Therefore, to be
able to follow this line of research, the technological development in
the observational side has to be accompanied by a similar development
in the lens modeling techniques in clusters of galaxies, to properly
account for the extra complexities associated with extended emission
and interferometric observations.


\begin{acknowledgements}

The authors wish to thank Sean Moran and Tadafumi Takata for providing
the HST and SUBARU images and source catalogs to compare with the
radio observations, and J.P. Kneib for providing the previous
{\scriptsize LENSTOOL} model on MS0451. We are also very grateful to
the anonymous referee for his/her useful comments and positive
feedback about the paper.

A.B.A would like to thank Christian Struve and Tom Oosterloo for
discussions and helpfull insights regarding the radio data reduction
and analysis, Oliver Czoske for helping with the alignment of the
different maps and Simona Vegetti for the useful discussions about the
lens model (which significantly improved the presentation of the model
results). A special mention is reserved for Edo Loenen and Ger de
Bruyn, for their crucial contribution in the analysis and
interpretation.

This work was supported by the European Community's Sixth Framework
Marie Curie Research Training Network Programme, Contract
No. MRTN-CT-2004-505183 ``ANGLES''. O.W. was funded by ANGLES and by
the Emmy-Noether-Programme of the Deutsche Forschungemeinschaft,
reference Wu 588/1-1. M.L. acknowledges the Centre National d' Etudes
Spatiales (CNES) for their support. The Dark Cosmology Centre is
funded by the Danish National Research Foundation.

\end{acknowledgements}

\bibliographystyle{aa}
\bibliography{12903}

\end{document}